\documentclass[11pt,aps,showpacs,nofootinbib,longbibliography]{revtex4-1}

\pdfoutput=1

\usepackage{ulem}

\usepackage{comment}
\usepackage[dvipsnames]{xcolor}
\usepackage{amssymb,amsmath,latexsym,mathrsfs}
\usepackage[USenglish,american]{babel}
\usepackage{bm}
\usepackage{amsfonts}
\usepackage{graphicx}
\usepackage{epstopdf}
\usepackage{hyperref}
\usepackage{array}
\usepackage[utf8]{inputenc}
\usepackage{soul}
\usepackage{color}
\usepackage{slashed}
\usepackage{csquotes}
\usepackage[T1]{fontenc}

\usepackage{footnote}

\everymath{\displaystyle} 

\usepackage{soul}

\newcommand{\beq}{\begin{equation}}
\newcommand{\eeq}{\end{equation}}
\newcommand{\beqa}{\begin{eqnarray}}
\newcommand{\eeqa}{\end{eqnarray}}
\newcommand{\bea}{\begin{eqnarray}}
\newcommand{\eea}{\end{eqnarray}}

\begin{document}

\title{
Oscillon spectroscopy
}

\author{Fabio van Dissel, Oriol Pujol\`as and Evangelos I. Sfakianakis} 
\email{\baselineskip 11pt Email addresses: fvdissel@ifae.es; pujolas@ifae.es; esfakianakis@ifae.es}

\affiliation{ Institut de F\'{\i}sica d'Altes Energies (IFAE)\\ 
 The Barcelona Institute of  Science and Technology (BIST)\\
 Campus UAB, 08193 Bellaterra (Barcelona) Spain
 }

\begin{abstract}

The sine-Gordon model in 3+1 dimensions is known to admit two oscillons of different energy and frequency but comparable lifetime.  
We {show that}  the oscillon spectrum 
includes more spherically symmetric ``states''. 
We identify new high-amplitude oscillons by allowing the field profile to have a number of nodes. 
For each number of nodes, we find 2 states with a comparable lifetime to the nodeless ones. %
Oscillons with nodes are, however, unstable to non-spherical perturbations and so their lifetime is significantly reduced. Interestingly, these states are seen to fragment into a collection of nodeless oscillons.
The heavy nodeless oscillon is quite remarkable: despite its energy it is stable against fragmentation. Moreover, it has considerably small oscillation frequency, meaning that it can be interpreted as a rather relativistic bound state.

\end{abstract}
\maketitle

\newpage 

\tableofcontents

\section{Introduction}
\label{sec:intro}

Oscillons are fascinating objects: they are non-perturbative bound states of bosonic fields held together by self-interactions but unprotected against decay. The tension between these two factors results in oscillons being evaporating quasi-bound states with a finite (but often considerably large) lifetime. 
Mathematically, this translates into the existence of quasi-attractor, localized and radiative solutions of the field equations, which  are long-lived. Usually, these do not admit closed form solutions.
Indeed, oscillons arise quite generically in simple scalar theories, including the ``sine-Gordon'' (SG) model in $3+1$ dimensions, which is of relevance for axions and which will be the main focus of this work. The lifetime of oscillons is model dependent, but even in common theories like SG it reaches  $\sim 10^{3}m^{-1}$. 

Since even in classical field theory oscillons must be constructed numerically (except in special cases, where analytic approximations are available), their ``discovery'' \cite{Bogolyubsky:1976nx,Bogolyubsky:1976sc,Gleiser:1993pt,Kolb:1993hw,Copeland:1995fq} was somewhat overlooked. 
More recently, they have gained substantial interest and several aspects have been further investigated, including their formation, longevity, their classical and quantum radiation as well as the model dependence of their behaviour, see e.g. \cite{Kasuya:2002zs,Amin:2010jq, Kawasaki:2015vga,Amin:2013ika,Ibe:2019vyo,Sfakianakis:2012bq,Olle:2020qqy,Hertzberg:2010yz,Fonseca:2019ypl,Levkov:2022egq} and \cite{Fodor:2019ftc} for a recent review.
Oscillons have been shown to emerge naturally in scenarios ranging from preheating to bubble collisions (see e.g. \cite{Amin:2011hj, Copeland:1995fq,  Gleiser:2009ys, Farhi:2005rz,  Graham:2006vy,  Zhou:2013tsa, Fodor:2008es, Hiramatsu:2020obh, VanDissel:2020umg,  Zhang:2020bec, Zhang:2020ntm, Antusch:2019qrr, Antusch:2017flz}), giving a gravitational wave signature \cite{Zhou:2013tsa,Hiramatsu:2020obh,Amin:2018xfe, Antusch:2017vga}. They also play a major role in dark matter \cite{Olle:2019kbo,Kawasaki:2019czd,Arvanitaki:2019rax,Olle:2020qqy,Cotner:2018vug,Kitajima:2020rpm,Kitajima:2021inh} 
and can be produced in topological defect networks \cite{Kolb:1993hw,Hindmarsh:2007jb,Braden:2015vza,Bond:2015zfa,Vaquero:2018tib,Gorghetto:2020qws,Kitajima:2022jzz,Blanco-Pillado:2020smt}. 
 
The aim of this work is to study one aspect that has received relatively little attention:  models that lead to oscillons typically admit not one but several oscillons. In fact, there is a rather well defined discrete {\it spectrum of oscillons} of increasing energy. 
One might expect that these higher energy oscillons must have a short lifetime.
Surprisingly enough, several of the ``excited'' oscillons have a lifetime comparable to  the lowest-lying oscillon.

\begin{figure}[t]
\centering
\includegraphics[width=.45\textwidth]{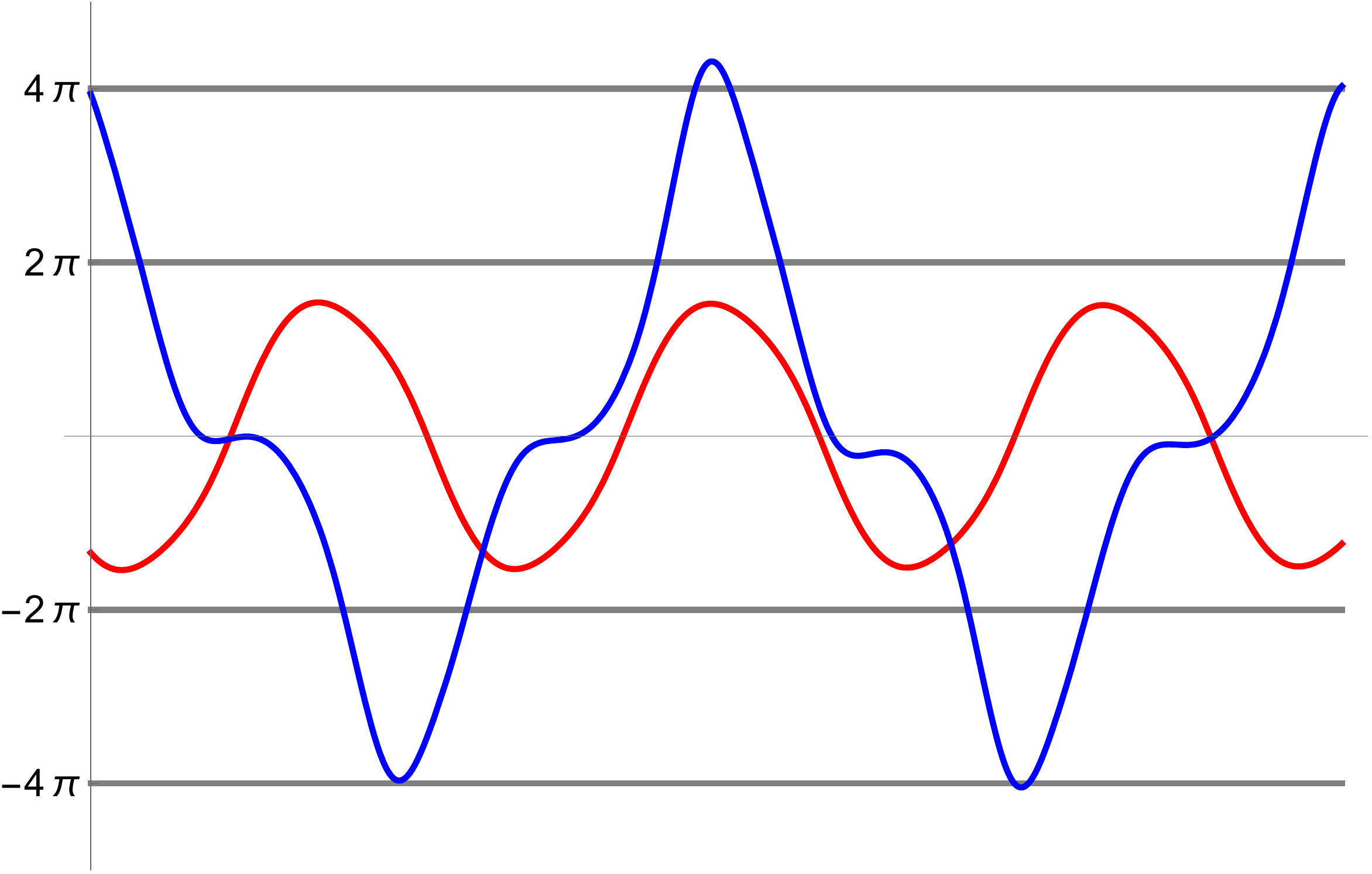}
\hspace{1cm}
\includegraphics[width=.45\textwidth]{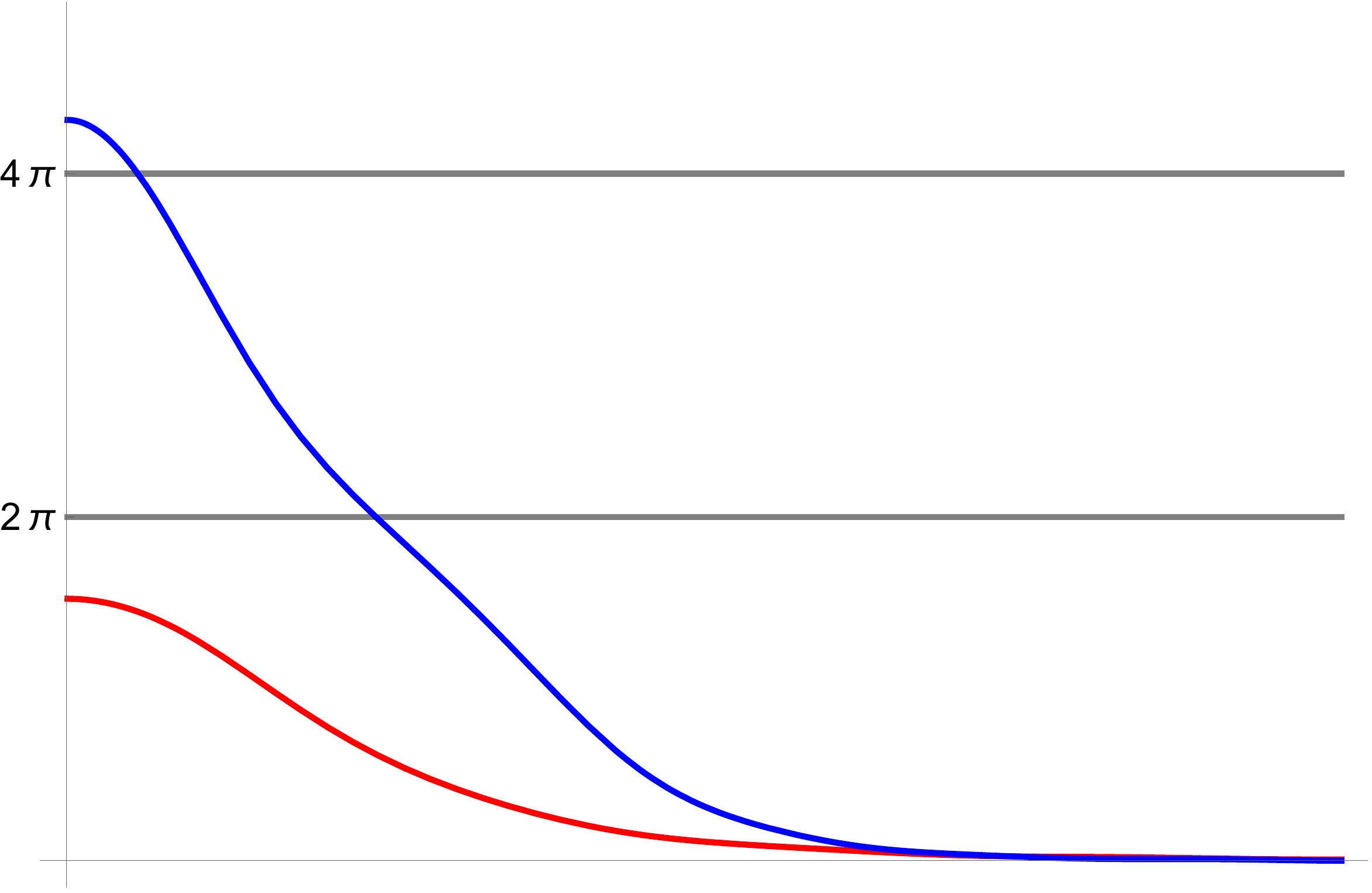}
\caption{Left: Amplitude of the field at the origin as a function of time, $\phi(t,0)$, for the two `fundamental' oscillons  of the  sine-Gordon model in $3+1$ dimensions. Right: Field amplitude as a function of the radius to the center. 
}
\label{fig:noNodes}
\end{figure}

For concreteness, we shall present and study these new excited oscillons in the SG model. The potential is
$$
V(\phi) = f^2 m^2 \left( 1- \cos\phi \right)
$$
where $\phi$ is the dimensionless field rescaled by the decay constant $f$, $m$ is the mass of $\phi$ quanta and we assume $m\ll f$. Ignoring $4\pi$ factors, $f$ is the UV cutoff of the theory, so $m\ll f$ translates into weak coupling.  
Our statements should extend qualitatively to other potentials, though quantitatively important differences may arise. 

{Oscillons can be also understood in  Quantum Field Theory as states with a high occupation number $N$ of the same single-particle state, see \cite{Dvali:2011aa,Dvali:2012gb,Dvali:2012en} and \cite{Olle:2020qqy}. For potentials with negative self-interaction, this state must be somehow preferred. However at weak coupling, $\lambda=m^2/f^2 \ll1$, the energy gain is only effective for $N \sim 1/\lambda$ \cite{Dvali:2011aa,Dvali:2012gb,Dvali:2012en}. This results into a large collective coupling, whence oscillons become non-perturbative. In this large occupation number limit, the mean-field description should be applicable, meaning that these states can be captured by solving classical field equations of motion. 
States with $N\sim f^2/m^2$ 
translate into classical field configurations with sizeable excursions $\phi\sim {\cal O}(1)$. The total mass of these configurations scales as $f^2/m$, with  oscillon-dependent prefactors\footnote{Since they oscillate with frequency close to $m$ and are localized in a radius of order $m^{-1}$, the highest gradients are typically at most $\sim f m$, well below the cutoff scale of the effective field theory, $\sim f^2$. }. Obtaining the oscillon spectrum thus becomes a non-perturbative problem -- there is no small expansion parameter. }

Most of the literature on  oscillons in the SG theory in $3+1$ dimensions refers to the lowest energy attractor oscillon. In this configuration the field at the origin $\phi(t,0)$ oscillates within the first period of the potential, that is 
{$|\phi|\lesssim 2\pi$,} as shown in Fig.~\ref{fig:noNodes}. Interestingly, the SG model is known to admit another oscillon \cite{Hormuzdiar:1998dn,Hormuzdiar:1999uz}, where  $\phi(t,0)$ oscillates in a larger range, reaching values of $\pm4\pi$, also shown in Fig.~\ref{fig:noNodes}. 
Let us emphasize a few (very) remarkable properties of the $\pm 4\pi$ attractor: 
\begin{itemize}
\item The total rest mass is about 4 times larger than the usual one. Yet, it is very stable -- as mentioned, its lifetime is comparable to the standard $\pm 2\pi$ oscillon.
\item As it is obvious from Fig.~\ref{fig:noNodes}, it contains a significant harmonic composition. The first harmonic (oscillating at $3\omega$) represents about $20\%$ of the profile $\phi(t,0)$. This harmonic has high enough frequency  to be  a radiation mode, however it is somehow ``trapped''  in the oscillon core. 
\item This solution is quite attractive to initial conditions with high enough (and localized enough) energy.
\item Depending on the initial conditions, the numerical evolution of the field equations starting near this attractor transitions adiabatically from the $\pm 4\pi$ attractor to the $\pm 2\pi$ one (thus collecting a total lifetime of more than $2000m^{-1}$).
\item Most remarkably, the oscillation frequency is quite low, $\omega \simeq 0.58 m$, to be compared with $\omega \simeq 0.92 m$ of the ``usual'' oscillon. This is quite extraordinary as the standard interpretation is that the difference between the mass of the  field and the frequency of the oscillon ($m-\omega$) represent the binding energy per quantum in the solution. This is, then, a significantly relativistic bound state. 
{Since, generally, the relativistic regime tends to make bound states significantly harder to understand, this system can provide useful insights.}
\end{itemize}

These observations open up many questions: Are there more oscillons (of similar lifetimes)? Is there any solution that explores even higher amplitudes? Is the oscillon spectrum discrete? Are they stable? Are there transitions among different oscillon states, and how do they proceed?

In this work we address these questions. We construct a sequence of new excited spherically symmetric oscillons with the defining property that they contain a given number of nodes. By ``node'' we mean a point in the radial coordinate where the field amplitude vanishes at all times.\footnote{Our notion of ``node'' and of ``excited oscillon'' thus differs from previous works, in particular from Ref.~\cite{Wang:2022rhk}. Oscillons with nodes were also discussed in Ref.~\cite{Fodor:2008es}, albeit in the small amplitude limit. } 
Oscillons with this kind of nodes are of course not expected to be attractors to generic initial conditions.\footnote{It is possible  that oscillons with nodes are attractors to initial conditions with nodes. Mapping the basin of attraction of these solutions
is beyond the scope of the present work.} 

However, the point of view of this work is to reveal the bound-state spectrum of the theory, or at least part of it. 
In some sense this is analogous to looking for the spectrum of resonances in  QCD. Even if most of them are unstable and/or  difficult to produce in practice, the spectrum of states with given properties (e.g.~spherical symmetry) is certainly well posed.

\begin{figure}[t]
\centering
\includegraphics[width=.8\textwidth]{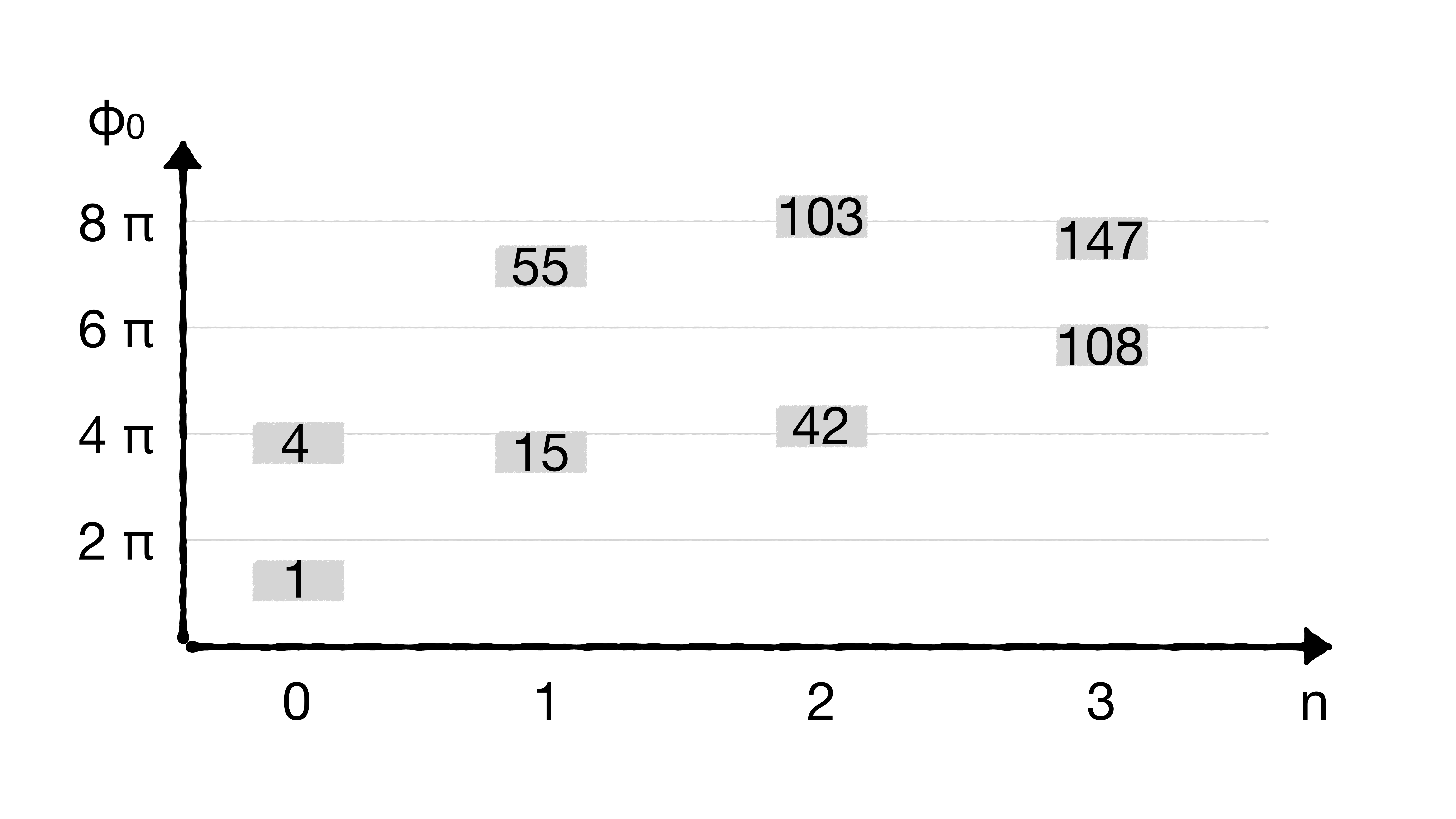}
\caption {The spectrum of long-lived spherically symmetric oscillon configurations. The horizontal axis corresponds to the number of nodes and the vertical axis shows the maximum field excursion at the center of the oscillon.
We stopped our exploration at oscillons with three nodes. The number in each box in the above plot shows the mass of each oscillon, with respect to the mass of the lowest-mass oscillon state, which is $\simeq 400 f^2/m$.}
\label{fig:exp}
\end{figure}

The spectrum of oscillon solutions that we find is shown in Fig.~\ref{fig:exp}, where we see that they span a large range in central amplitude and total energy (rest mass).
We do not go beyond states with 3 nodes, but in principle the sequence continues, perhaps indefinitely.
Each point corresponds to a different oscillon that keeps the number of nodes for a significant time, which we demand to be comparable to the lifetime of the lowest lying oscillon. 
{ Table \ref{tab1} summarizes the basic properties of these states. } The error bars indicate the range in amplitude $\phi_0$ and frequency $\omega$ that is explored by each oscillon {during its lifetime. }

Quite interestingly, we find that i) the spectrum of oscillons is discrete; ii) the central field value $\phi_0$  takes on specific values, which are close to multiples of $2\pi$; iii) oscillons with large amplitude $\phi_0\gtrsim 4\pi$ exist, at the expense of introducing nodes; iv) there are more examples of very relativistic oscillons (with $\omega$ significantly below $m$).

The spectrum shown in  Fig.~\ref{fig:exp} is obtained using a  combination of  analytic and numerical methods. 
Introducing a single frequency ansatz, we reduce the equation of motion to an effective one-dimensional problem. This provides a guess which is used as the initial condition for simulating an oscillon with nodes but which does not correspond to the actual oscillon profile. The field then quickly relaxes to an excited oscillon with nodes 
and high harmonic content, meaning that these ``guess'' initial conditions fall inside the basin of attraction  of these oscillons.

{Two remarks are in order. First, it is clear that the sine-Gordon {model} has a larger class of spherically symmetric solutions, in the form of spherical domain walls (kinks) with an initial radius $R_0\gg1/m$. With perfect spherical symmetry, after a time of order $R_0$ they reach small sizes and bounce back emitting radiation and re-bounce a number of times \cite{Widrow:1989vj}. In a way, then, these solutions are similar to  oscillons. (In fact, some of them end up trapped in oscillons) Of course, there is a continuum of solutions of this form and their lifetime is large, but we discard them as we are interested in states with well defined properties (number of nodes, radius, etc) staying constant or at most changing adiabatically throughout the lifetime.}

{Second, the spectrum in Fig.~\ref{fig:exp} is selected by imposing both a long lifetime and spherical symmetry. Relaxing the symmetry can of course impact on the actual lifetime of these states -- as it clearly also affects the spectrum. (With less symmetry, multi-oscillon solutions could exist.)  In Section~\ref{sec:conclusions} we 
study how allowing for aspherical modes affects these states. As it turns out, the lifetimes of oscillons are affected, but the heavy nodeless one is not.  
Quite remarkably, we
see the breakup of an  ``excited'' oscillon to a (large) collection of nodeless ones. }

The current work is organized as follows. In Section~\ref{sec:model} we describe the model and provide analytical constructions for oscillons, using both as single-frequency ansatz as well as considering contribution from higher harmoncis.  We provide oscillon solutions with up to $3$ nodes and examine their properties. Section~\ref{sec:sphericalevolution} contains an extensive set of numerical simulations, where the evolution of oscillons is computed  in the spherical ansatz. In Section~\ref{sec:transitions} we discuss possible transitions between oscillon states and show three-dimensional simulations.  We conclude and provide suggestions for future work in Section~\ref{sec:conclusions}.

\section{Quasi-breather approximation}
\label{sec:model}

The action we  use is
\beq
\label{eq:action}
\Phi =f^2\,\int d^3x\, dt\,  \left [
{1\over 2}\partial_\mu \phi \partial^\mu\phi - m^2 (1-\cos\phi)
\right ]
\, ,
\eeq
where we have pulled out an overall factor given by the decay constant $f$ so that the field $\phi$ is dimensionless. 
Finding the spherically symmetric oscillons in this theory reduces then to look for the configurations $\phi(r,t)$ that solve the equation of motion
\beq
{\partial^2 \phi \over \partial t^2}- {\partial^2\phi\over\partial r^2} - {2\over r}{\partial\phi\over\partial r} + m^2\, \sin\phi=0~,
\label{eq:SGeom}
\eeq
with outgoing radiation boundary conditions. 

\begin{figure}[h]
\centering\includegraphics[width=.7\textwidth]{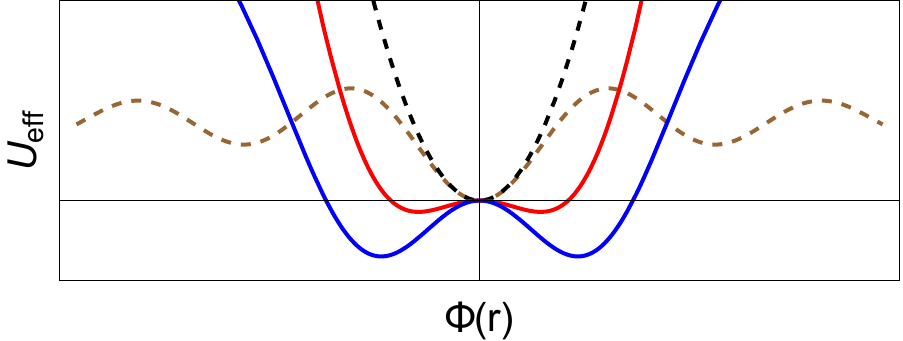}
\caption{ The effective potential for the single-frequency ansatz for $\omega=0.5,0.8$ (blue and red respectively). The dashed lines correspond to ${1\over 4} \Phi^2$ (black) and $1-J_0(\Phi(r))$  (brown).
}
\label{fig:effectivepotential}
\end{figure}

The construction of oscillons starts by treating them as ``quasi-breather'', a periodic localized configuration, and optimize the radial profile and the frequency to identify slowly radiating configurations. One starts with a single-frequency ansatz 
\beq\label{eq:singlew}
\phi(r,t) = \Phi(r) \sin(\omega t)
\eeq
with $\omega <m$ for the oscillon solution to exist. 
By inserting this ansatz into the action and integrating over time over a period $2\pi/\omega$, we arrive at the effective action 
\beq
S_{\rm eff} = f^2\int d^3x
\left \{
{1\over 4}\omega^2 \Phi^2 + {1\over 4}[\nabla \Phi]^2
-[1-J_0(\Phi)]\right \}
\eeq
which in turn leads to the equation of motion for the amplitude $\Phi(r)$
\beq
\label{eq:profile}
{d^2\Phi \over dr^2}+ \frac{2}{r} {d\Phi\over dr} + \omega^2 \Phi - 2 J_1(\Phi) = 0~,
\eeq
where $J_n$ are Bessel functions of the first kind.
The solution $\Phi(r)$ needs to be regular (zero derivative) at the origin  and vanish at spatial infinity.
This equation of motion has an intuitive mechanical analogue; that of a point particle moving in a one-dimensional potential well with potential $U_{\rm eff} = {1\over 4}\omega^2 \Phi^2 -[1-J_0(\Phi)] $. In this analogy $r$ is the time coordinate and the  term $-{2\over r} \dot \phi$ describes friction.
The boundary conditions  for finding oscillon solutions are equivalent to a ball starting at some large value $\Phi(r=0)$ and rolling towards the origin, reaching it in infinite ``time'' $r\to \infty$. 
Fig.~\ref{fig:effectivepotential} shows the effective potential for different values of the oscillon frequency $\omega$. 

The simple numerical procedure for finding oscillon solutions includes choosing a frequency $\omega<1$ and searching for the proper value of $\Phi(r)=0$ that leads to $d_r \Phi(r\to\infty) =0$, corresponding to a solution where the point particle is released at rest from some amplitude and asymptotically reaches the origin.

This of course neglects the case of the ball overshooting the point at the origin and probing negative values of $\Phi$, later to return and reach the origin in infinite time from the left. 
Each time the ball overshoots the local maximum at the origin, the corresponding oscillon solution acquires a node. This construction leads   to a two-parameter family of solutions, which is continuous in $\omega<1$ and discrete in the number of nodes, ranging from $0$ to infinity (in principle).

Since the above construction (and any oscillon) does not lead to exact solutions of the equation of motion of the full system, oscillons in general will also have a radiating tail at higher harmonics. By introducing only the  {third harmonic}\footnote{We refer to the mode oscillating at the fundamental frequency $\omega$ as the  first {or fundamental} harmonic. {For notational clarity, we must stress the difference between the {terminology} ``fundamental oscillon'', having no nodes as shown in Fig.~\ref{fig:exp} and the fundamental frequency / harmonic, which we define here.}}, the ansatz becomes 
\beq
\phi(r,t) = \Phi(r)\sin(\omega t) + \Phi_3(r)\sin(3\omega t)~. 
\eeq
Assuming $\Phi_3\ll \Phi$, it is straightforward to see whether there are special points in solution space, where the third harmonic is non-radiating. All we need to do is solve the  equation
\beq
{d^2 \Phi_3(r)\over dr^2} +{2\over r}{ d \Phi_3(r)\over dr} + (3\omega)^2 \Phi_3(r) - J_0(\Phi(r)) \Phi_3(r)=
2J_3(\Phi(r))
\label{eq:S3linear}
\eeq
{where the emergence of Bessel functions of different order is described in Appendix~\ref{app:multifrequency}.}
The inhomogeneous solution to the above equation can be found through the Greens's function method
\beq
\Phi_3(r) = \int_0^r {\cal G}(r,r') 2J_3(\Phi(r'))dr'
\eeq
The condition for a localized third harmonic solution is simply the vanishing of the above integral for $r\to\infty$. This is also derived in Ref.~\cite{Cyncynates:2021rtf} using arguments based on destructive interference.

When the amplitude of the main harmonic $\Phi(r)$ grows larger, so do in general the corresponding higher harmonics. This means that after some point, the linear approximation of Eq.~\eqref{eq:S3linear} fails to capture the dynamics of the system, since the third harmonic will be large enough to back-react onto the {first harmonic}. Ref.~\cite{Cyncynates:2021rtf} proposed a quasi-breather construction that allows one to include the contribution of higher harmonics, taking their mutual interactions into account. In this formalism, the oscillon is described (taking only the first and third harmonics into account) as $\phi \simeq \Phi_1 \sin(\omega t) + \Phi_3 \sin(3\omega t) +c_3\cos(3\omega t)$, 
 where the term $c_3$ is added in order to allow for outgoing radiation at $r\to\infty$, which wouldn't be possible with an expansion solely in terms of sines or cosines. 
The equations that one needs to solve are
\begin{eqnarray}
&&{d^2\Phi_1\over dr^2 } + {2\over r} {d\Phi_1\over dr} + \omega^2\Phi_1 + f\left(\Phi_1,\Phi_3\right )=0 
\\
&&{d^2\Phi_3\over dr^2 } + {2\over r}{d\Phi_3\over dr} + 9\omega^2\Phi_3 + g\left(\Phi_1,\Phi_3\right )=0 
\\
&&{d^2c_1\over dr^2 } + {2\over r} {d c_3\over dr} + 9\omega^2c_1 + h\left(\Phi_1,\Phi_3\right )c_3=0 
\end{eqnarray}
where $f,g,h$ are functions of $\Phi_1$ and $\Phi_3$ and we assumed that $c_3$ can still be treated in the linearized approximation. The derivation and exact form of these equations is given in Appendix~\ref{app:multifrequency}.

\subsection{Fundamental Oscillons}

\begin{figure}[h]
\includegraphics[width=.32\textwidth]{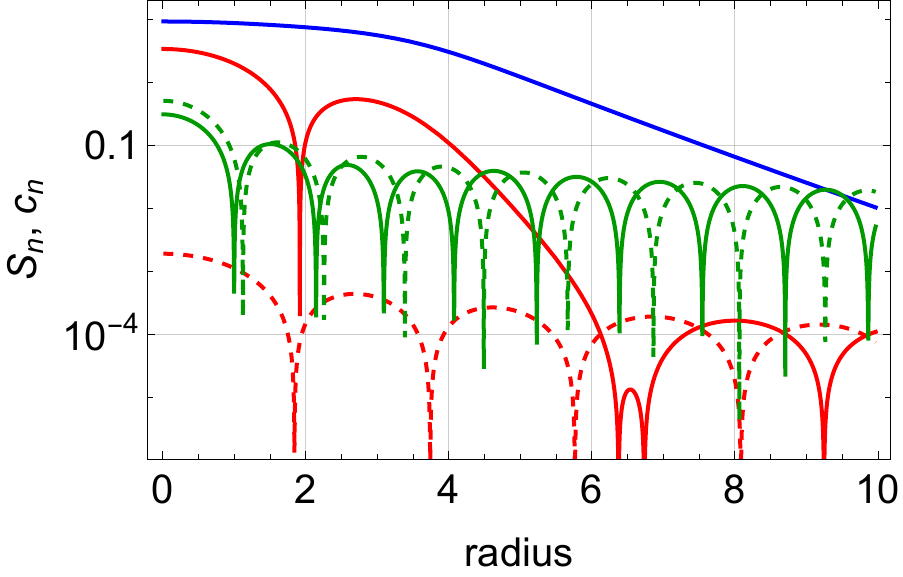}
\includegraphics[width=.32\textwidth]{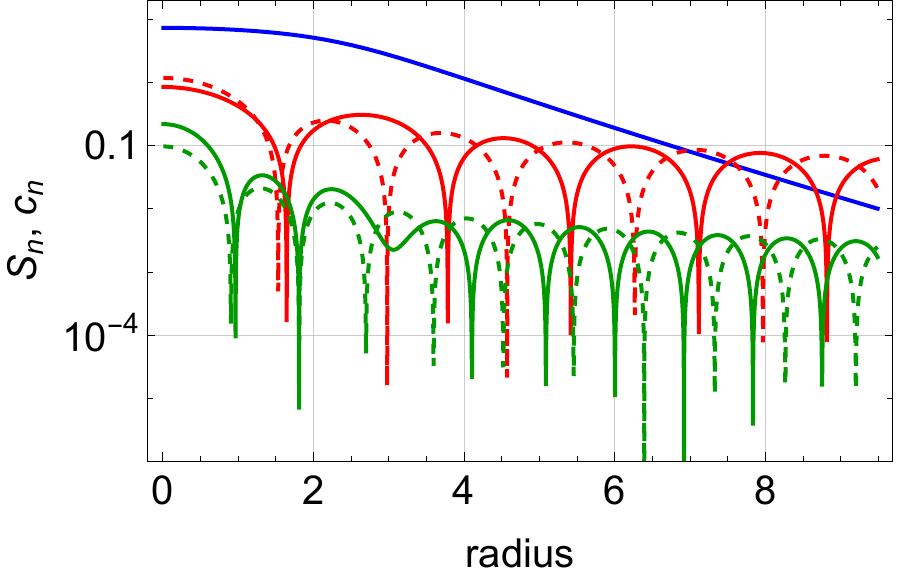}
\includegraphics[width=.32\textwidth]{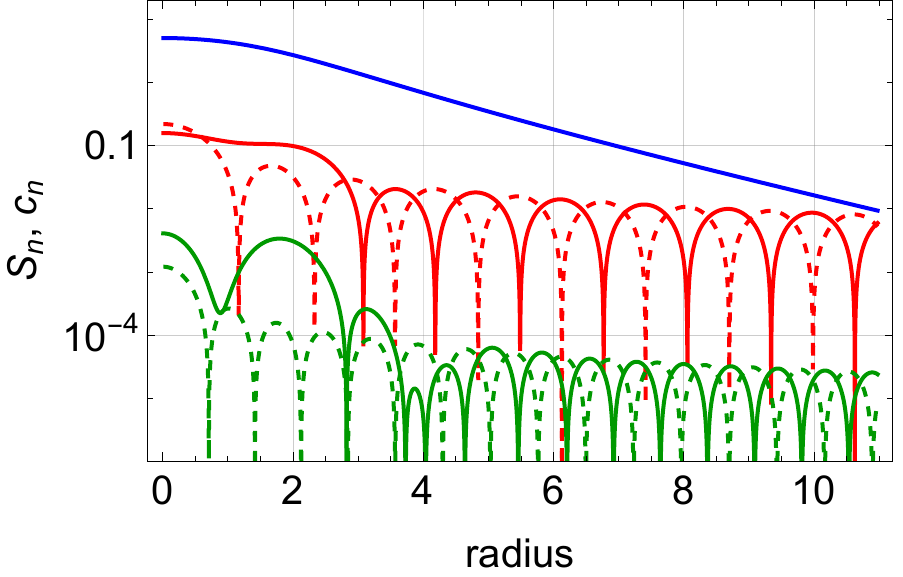}
\caption{Oscillon solutions for $\omega = 0.56, 0.7,0.92 $ (left to right). The first, third harmonic and fifth harmonics are shown in blue, red and green respectively (solid for sines and dashed for cosines) }
\label{fig:node0analytics}
\end{figure}

We start by constructing and studying the ``fundamental'' oscillon solutions, which in this context means oscillons without nodes in their spatial profile.  Fig.~\ref{fig:node0analytics} shows the two long-lived oscillon states that emerge for $\omega /m\simeq   0.56 , {0.92}$  (left and right panels), along with a ``random'' solution at $\omega = 0.7$. It is clear that the solution shown in the middle panel has a much larger radiating tail, leading to a faster loss of energy. This drives the oscillon towards the long-lived solution at $\omega \simeq 0.92$. 

Fig.~\ref{fig:node0analytics} contains interesting information about the structure of the fundamental oscillons in the three dimensional sine-Gordon equation. We first notice the long-distance behavior: The first harmonic is exponentially decaying, while the third and fifth ones are oscillating, both in the sine and in the cosine. However, the two terms in each harmonic have a constant shift of $\pi/2$, leading to a traveling wave towards infinity. Thus this regime represents the radiation emanating from the oscillon.

We see that for the longer lived oscillons ($\omega/m=0.56,{0.92}$) the radiation tail is about one order of magnitude smaller than a ``random'' solution with $\omega/m = 0.7$. Furthermore, for the oscillon with $\omega/m=0.56$, the radiation of the third harmonic is several orders of magnitude suppressed and the decay is controlled by the fifth harmonic. For the case of  {$\omega/m=0.92$} the opposite occurs and the fifth harmonic is vastly subdominant. 

From the sequence of panels in Fig.~\ref{fig:node0analytics} we see how the third harmonic evolves as a function of the frequency and the oscillon height. The height of the first harmonic becomes larger for smaller frequencies and this leads to a more significant excitation of the third harmonic. For $\omega=0.56$ the third harmonic is non-perturbatively large close to the oscillon core. In this range of frequencies, the third harmonic is largely confined, as is evident from the significant difference between the sine and cosine terms near the core as well as from the difference between the value of the third harmonic near and far from the core of the oscillon. Contrary to that, the fifth harmonic is always perturbative (much smaller than the {first} harmonic). Furthermore, the structure of the three first harmonics elucidates the difference between near and far regions of the oscillon. In the far region, the two terms in each harmonic constituting the radiating tail have a phase difference of $\pi/2$, whereas near the core they are in phase. 
For $\omega=0.7$ we see that both higher harmonics change behavior near $r=3$. Interestingly for the long-lived oscillon of $\omega/m=0.56$ the third harmonic becomes radiating for $r\gtrsim6$ and the fifth harmonic for $r\gtrsim 4$, meaning that the core and tail of the oscillon can be perceived slightly different for different harmonics.

\begin{figure}[h]
\centering
\includegraphics[width=.45\textwidth]{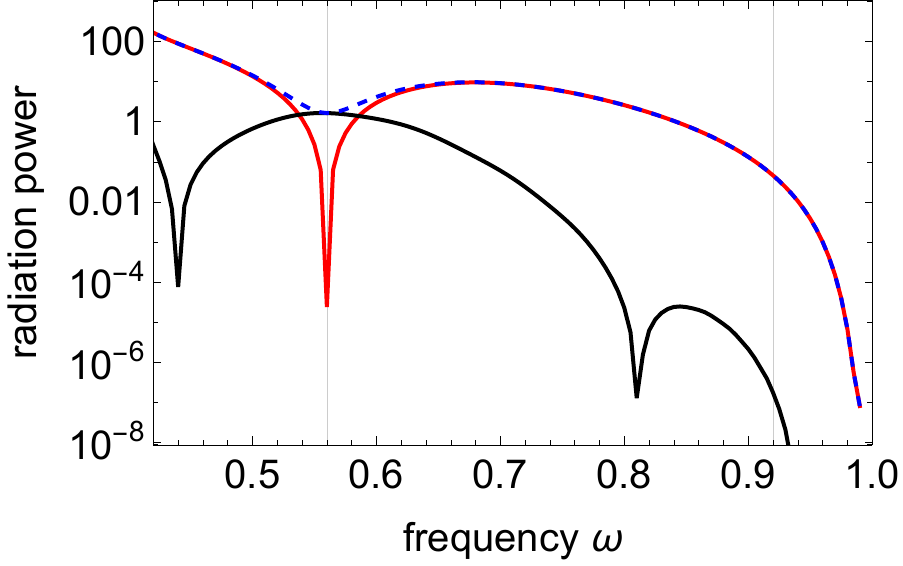}
\includegraphics[width=.46\textwidth]{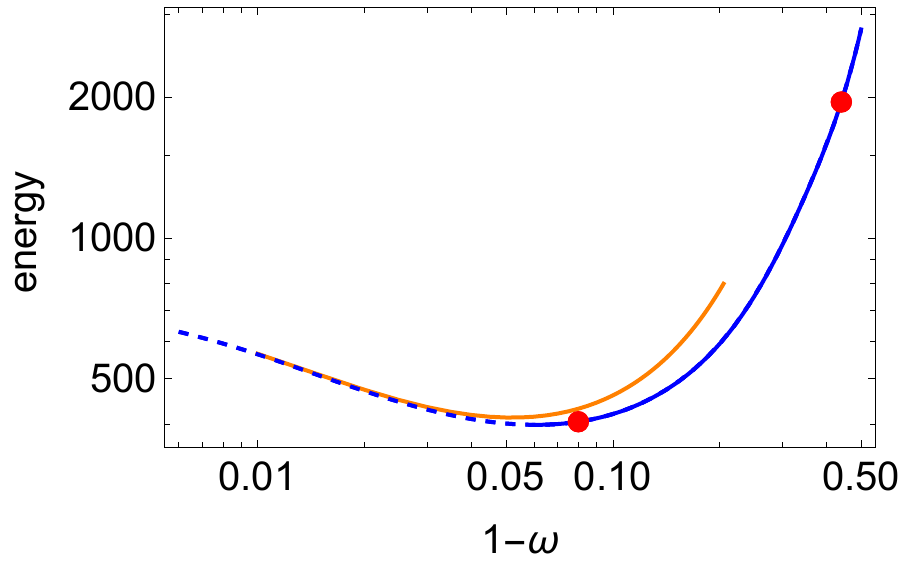}
\caption{{\it Left:} The radiated power of the node-less oscillons in the third and fifth harmonic (red and black respectively), along with the total radiated power (blue-dashed).
{\it Right:} The oscillon energy as a function of frequency (blue). The two red dots correspond to the two long-lived states.  The solid (dashed) part corresponds to states which are stable (unstable) with respect to long-wavelength perturbations. The orange curve shows the  particle number $N$ {(defined as $E_{osc}/\omega$)}, around the point where the stability behavior changes.}
\label{fig:node0analyticsdecay}
\end{figure}

Since oscillons are not exact solutions of the equation of motion and they contain small radiating tails (even suppressed ones), one can define the rate at which energy is expelled from the oscillon towards infinity. The energy loss is directly related to the change in oscillon energy per unit time $\Gamma_E(\omega) = - \dot E(\omega)$ and is shown in Fig.~\ref{fig:node0analyticsdecay}. We see that the third harmonic has a dip at $\omega\simeq 0.56$. By ``zooming'' in close to this dip, we can see the third harmonic vanish at large distances, within the numerical accuracy limits of our calculation. At this point, the energy loss is dominated by the fifth harmonic. Thus $\omega\simeq 0.56$ defines a local minimum in energy loss, where a long-lived oscillon state is expected to exist. 
The behavior of the energy loss at high frequencies, close to $\omega=1$ is somewhat different. The energy loss there decreases monotonically, without providing a clear local minimum. 
As the oscillon loses energy, its frequency increases. However, this cannot continue arbitrarily, since the energy function has a minimum for $\omega\simeq 0.93$, meaning that after this point the oscillon cannot lose energy and evolve adiabatically, leading to its sudden decay. This has been coined ``energetic death" \cite{Cyncynates:2021rtf}.

\begin{figure}[h]
\centering
\includegraphics[width=.45\textwidth]{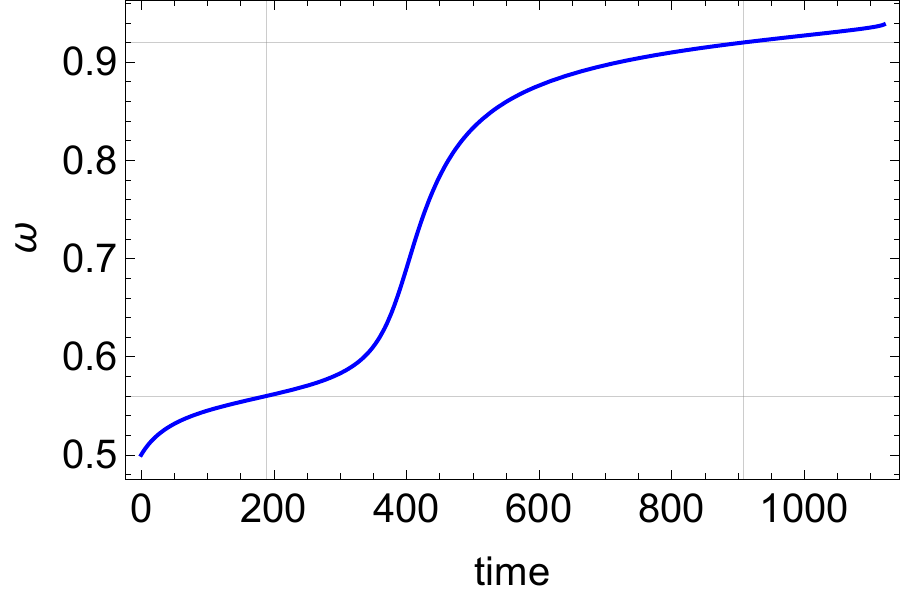}
\includegraphics[width=.48\textwidth]{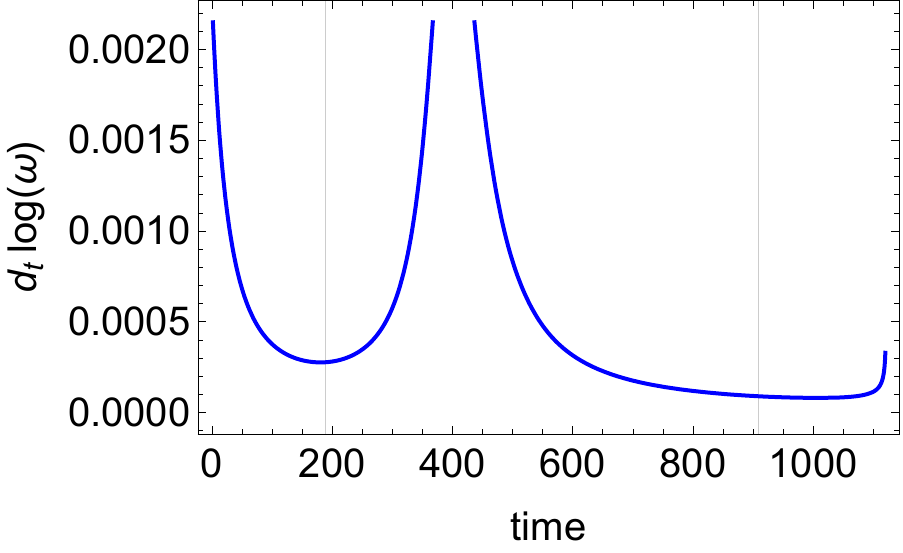}
\caption{{\it Left:} The evolution of the instantaneous frequency $\omega(t)$ as a function of time. We can clearly see two plateaus, corresponding to well-defined oscillons.
{\it Right:} The relative change in frequency $\dot\omega/\omega$, clearly pointing to the existence of two clear oscillon states, at $\omega\simeq 0.56$ and $\omega\simeq 0.92$.}
\label{fig:omegavstime}
\end{figure}

We can look for oscillon  configurations in the space of instantaneous solutions, which are able to retain their global properties over a substantial amount of time. Simply put, we can look for solutions, whose frequency does not change over a long period of time.
From the quantities derived in the semi-analytic oscillon solution, we can compute the the evolution of the instantaneous frequency of the system by solving the following equation numerically
\beq
\int_{\omega_0}^\omega\frac{dE/d\omega'}{\Gamma_E(\omega')} d\omega' = -\int_{t_0}^t dt' 
\label{eq:omegavst}
\eeq
Fig.~\ref{fig:omegavstime} shows two clear minima of $\dot\omega/\omega$, the relative evolution of the instantaneous frequency. One is at $\omega\simeq 0.56$, as expected, and the other can be found at $\omega\simeq0.92$, leading to an concrete definition of long-lived oscillon states.

\subsection{Excited Oscillons}

We now move to  oscillon solutions with $1, 2$ and $3$ nodes. 
By performing the same shooting process, in order to find oscillon solutions, but starting with a slightly larger amplitude at $r\simeq 0$, we discover a family of oscillon solutions where the shape of the {first (fundamental)} harmonic exhibits a node, a point in space where the amplitude vanishes. We should note that, within the validity of our construction method, the total amplitude does not vanish at all times, due to the fact that the higher harmonics do not vanish at the same point. However, they are vastly subdominant (one order of magnitude or more).  We will thus use the characterization of nodes based on the  {first} harmonic and keep in mind that in reality they are ``approximate nodes''.

\begin{figure}[h]
\includegraphics[width=.32\textwidth]{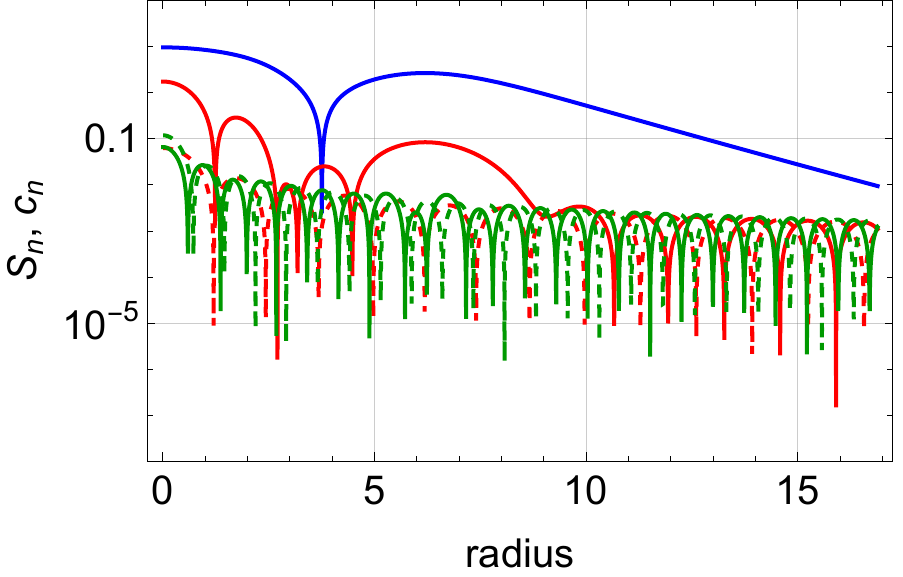}
\includegraphics[width=.32\textwidth]{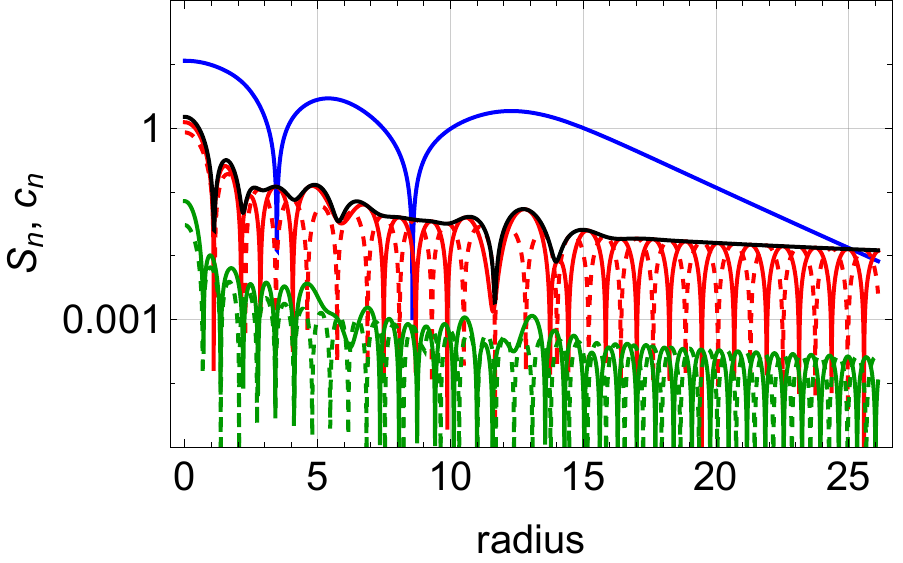}
\includegraphics[width=.32\textwidth]{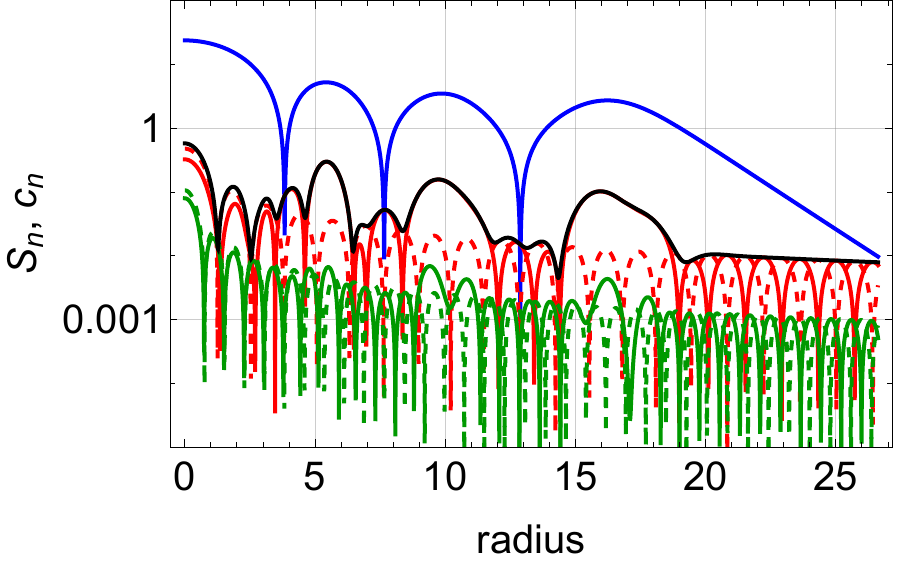}
\\
\includegraphics[width=.32\textwidth]{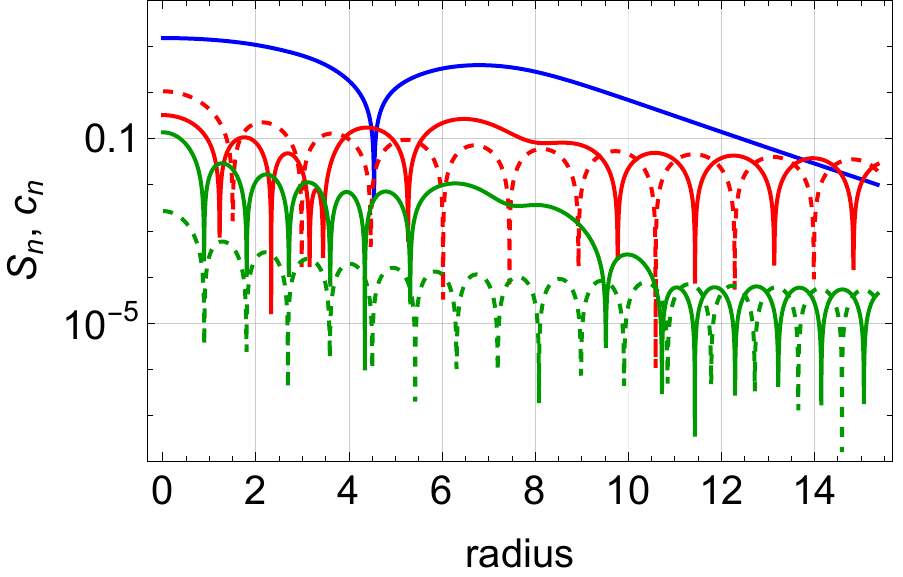}
\includegraphics[width=.32\textwidth]{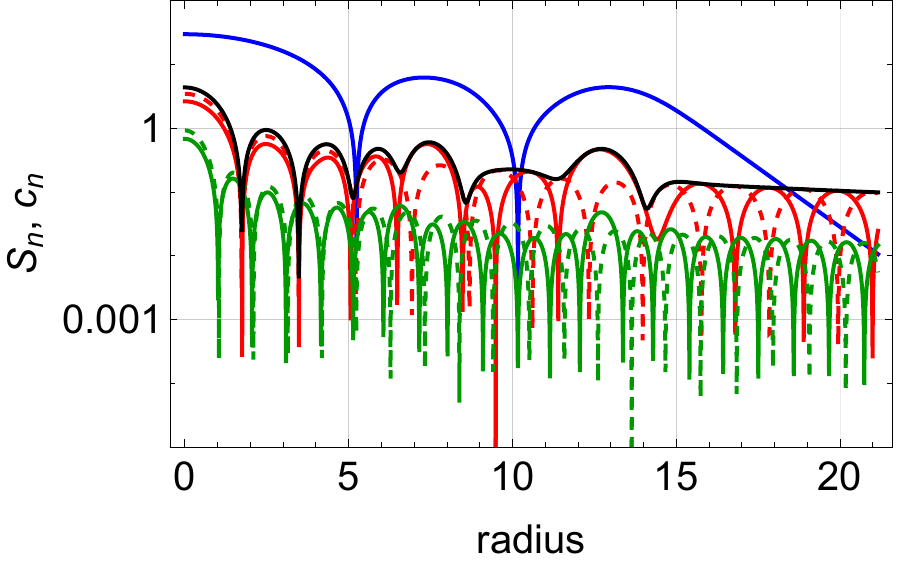}
\includegraphics[width=.32\textwidth]{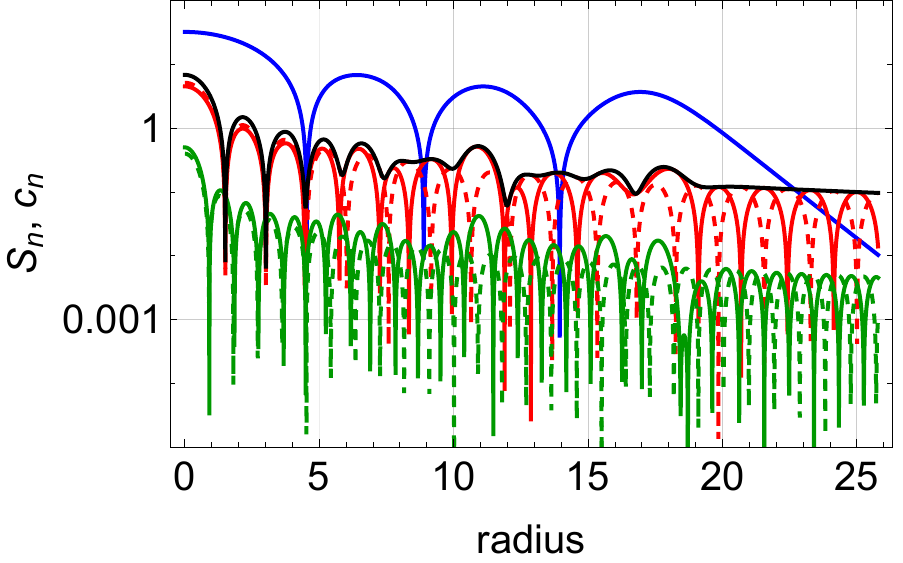}
\caption{Oscillon solutions for $n=1, 2, 3$ nodes (left to right). The upper panels correspond to long-lived solutions, $\omega =  0.86,0.92,0.82 $ respectively.
The lower panels correspond to ``random'' solutions $\omega = 0.7,0.6,0.7 $ respectively. The first, third harmonic and fifth harmonics are shown in blue, red and green respectively (solid for sines and dashed for cosines). The black curve corresponds to the the sum of squares of the corresponding sine and cosine terms of one higher harmonic.}
\label{fig:node1analytics}
\end{figure}

Fig.~\ref{fig:node1analytics} shows the structure of  oscillons with $1$, $2$ and $3$ nodes for two different frequencies each. One corresponds to a solution where the decay rate exhibits a local minimum, leading to long-lived state, whereas the other frequency does not possess such features.
We see a similar behavior as the one shown in Fig.~\ref{fig:node0analytics}, with the outgoing radiation being suppressed for certain values of the frequency. Interestingly, the higher harmonics are in general suppressed at the nodes. In some cases, (some) higher harmonics show an outgoing radiation-like behavior, meaning that the sine and cosine terms are exactly out of phase. In order to capture this, we simply add the squares of the two contributions to the harmonic in question (black lines in Fig.~\ref{fig:node1analytics}). This should not have features for outgoing radiation, or whenever the sine and cosine terms oscillate out of phase.

\section{Numerical Evolution in Spherical Symmetry}
\label{sec:sphericalevolution}

In this section we take an ``experimental" point of view and use numerical methods to study the oscillon states of the sine-Gordon model. We employ a simple central difference algorithm to study the time evolution of these states imposing spherical symmetry and implementing absorbing boundary conditions (we will discuss the nontrivial question of non-spherical breakup of oscillon states in Section~\ref{sec:transitions}). We checked that our results are robust to a variety of integration schemes and resolutions. The main points of this section can be summarized as follows:
\begin{itemize}
    \item We take a different viewpoint here than the one provided in Section~\ref{sec:model}. There, the quasi-breather picture gives a continuously connected spectrum of oscillon-like configurations. We'll show that, although some aspects of the dynamics of oscillon states can be captured in this way, generically other effects dominate the evolution of the system. We thus take a more restrictive approach to what we call a ``state'': {\it an oscillating field configuration  
    that is able to retain most of its properties during its lifetime}. Although these properties can be understood in the quasi-breather framework, our current definition naturally leads to the emergence of a discrete spectrum of states.
    \item The single-frequency ansatz defines the initial conditions of our ``experiment", labeled by frequency (continuous) and number of nodes in the spatial profile (discrete). We observe that not all of these initial conditions relax to our definition of a state. Using the quasi-breather picture we can semi-analytically predict where these states exist, finding good agreement with experiment. However, this framework doesn't seem 
    appropriate to predict the overall lifetime of the states, which is dominated by additional instabilities. This effect is especially prevalent for states that carry nodes. Lifetimes discussed in this section are thus directly extracted from  numerical simulations.
    \item As instabilities play an important role for the dynamics of these oscillons, the overall lifetime of the states is tied to the exact initial conditions that we choose (starting with a large perturbation around the exact state leads to an earlier onset of decay). When assigning lifetimes, we include uncertainties to account for this. 
    \item We find a discrete spectrum of long lived oscillon states. We characterize states as ``long lived'', when they have a lifetime comparable ($\gtrsim 50\%$) to the fundamental (lowest energy) state. By adding more and more nodes to the initial conditions a seemingly boundless spectrum emerges, with ever increasing central amplitude and total mass $M$ (rest energy). We explicitly checked that states can be found with up to 3 nodes in the spatial profile, and there is no a priori reason to suspect that this process can't continue up to an arbitrary number of nodes. It is at this point not clear if this is a property of the sine-Gordon model with its infinite degenerate minima, and this question will be addressed in future work. Our results are summarized in Table~\ref{tab1}.
\end{itemize}

\begin{table}[h]
\begin{center}
\begin{tabular}{| c | c | c | c | c | c |}
\hline
 ~$n$~ & $\omega$   $(\cdot m)$& $\Phi_0$ $(\div 2 \pi)$ & $\tau_{sp}$ $(\cdot 10^2 m^{-1})$  &  $M$ $(\cdot 10^2 f^2/m)$& $ M \tau_{sp}$ $(\cdot 10^6\,f^2/m^2)$ \\ 
 \hline \hline
 $0$ & $0.58(5)$ & $1.92(9)$ & $9.5(5)$ & $17.0(8)$ & $1.6(2)$ \\ 
 \hline
  $0$ & $0.92(1)$ & $0.68(5)$ & $7.0(5)$ & $3.8(2)$ & $0.27(3)$
 \\ 
 \hline
  $1$ & $0.56(4)$ & $3.6(4)$ & $6.0(5)$ & $210(40)$ & $13(3)$
   \\ 
 \hline
  $1$ & $0.86(2)$ & $1.8(2)$ & $6(1)$ & $56(1)$ & $3.4(6)$
   \\ 
 \hline
  $2$ & $0.70(5)$ & $4.1(8)$ & $3.5(1.0)$ & $392(7)$ & $13(4)$
   \\
   \hline
  $2$ & $0.92(1)$ & $2.1(1)$ & $5(1)$ & $160(10)$ & $8(2)$
   \\ 
 \hline
  $3$ & $0.82(2)$ & $3.8(2)$ & $7(1)$ & $557(8)$ & $39(6)$
  \\
  \hline
  $3$ & $0.91(2)$ & $2.8(1)$ & $5.5(5)$ & $410(20)$ & $23(3)$
  \\
 \hline
\end{tabular}
\end{center}
\caption{{Properties of long lived, spherically symmetric oscillon solutions: node number ($n$), frequency ($\omega$), field amplitude at the center ($\Phi_0$), lifetime ($\tau_{\rm sp}$), mass $M$ and narrowness ($M\tau_{\rm sp}$). We list only oscillon states which retain well defined properties (e.g., the number of nodes) for a lifetime comparable to the lowest mass oscillon. 
 The  variance  is indicated in parenthesis notation, e.g. $ \omega = 0.58(5) \Rightarrow 0.53 \le \omega\le  0.63$. }The lifetime is written as $\tau_{sp}$ emphasizing the fact that these are lifetimes obtained in spherical symmetry. It seems that the lifetime is significantly shortened if aspherical perturbations are included.}
\label{tab1}
\end{table}

\subsection{Experimental Setup}
\subsubsection{Initial Conditions 
and radiated power}
\begin{figure}[h]
    \centering
    \includegraphics[width=0.6\textwidth]{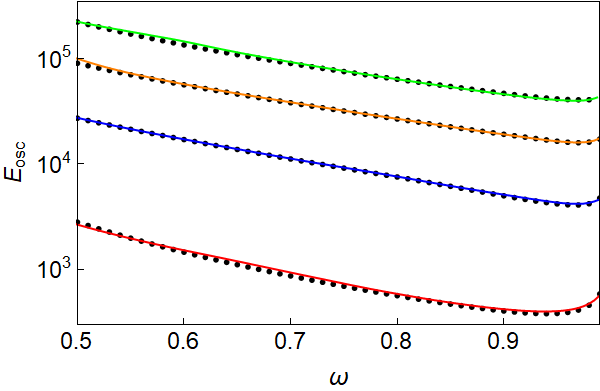}
    \caption{Comparison of the energy (or rest mass) obtained analytically with the full harmonic content in the quasi-breather formalism (color) and the single-frequency ansatz (black dotted) of the oscillon-like configurations of the sine-Gordon model, for different number of nodes in the spatial profile: $0$ (red), $1$ (blue), $2$ (orange) and $3$  (green). The difference is negligible.
    }
    \label{fig:energyoscillons}
\end{figure}

The results of our numerical experiment are entirely determined by 
the evolution of the system through the equation of motion, Eq.~\eqref{eq:SGeom}, 
as well as by
the initial and boundary conditions. We impose spherical symmetry on the system $\phi(\Vec{x}, t) \to \phi(r, t)$, and implement an absorbing boundary condition at $r \rightarrow \infty$ and a Neumann boundary condition at $r = 0$;  $\partial_r \phi(0, t) = 0$. 
This uniquely defines the problem, once the initial conditions are specified.
To probe the space of possible oscillon states we use the single-frequency ansatz to guide our choice of initial conditions.
This was described in Section~\ref{sec:model} and numerically calls for the solution of Eq.~\eqref{eq:profile}
with boundary conditions $\partial_r\Phi|_{r=0} = 0$ and $\Phi(r\to \infty) = 0$.
In practice, the boundary condition at infinity is substituted for the boundary condition at a finite but large radius, where the equation of motion can be approximated as
\beq
{d^2 \Phi \over dr^2} + {2\over r}{d\Phi\over dr} + (\omega^2-m^2)\Phi\simeq 0
\eeq
which leads to the trivial solution
$\Phi \sim {1\over r}e^{-\sqrt{m^2-\omega^2}\,r}$.
At fixed frequency, this procedure defines a discrete spectrum of initial conditions for our setup, where the solutions are labeled by the amount of nodes in the spatial profile $\Phi(r)$. We then choose initial conditions to satisfy the single-frequency ansatz, with $\phi(r, 0) = \Phi(r)$ and $\partial_t{\phi(r, t)}|_{t=0} = 0$. In practice, true oscillons have a non-negligible contribution from higher harmonics.
\\ \\
The presence of higher harmonics in the full solution has three effects which can be studied numerically as well as analytically within the quasi-breather picture outlined in Sec.~\ref{sec:model}
\begin{itemize}
    \item There is a significant deviation of the spatial and temporal profile of the oscillon states with respect to the single-frequency ansatz. This effect is especially noticeable in large amplitude oscillons.
    \item The oscillon radiates and thus loses energy through unbounded higher harmonics.
    \item As the oscillon loses energy it is forced to adiabatically change its frequency and find solutions that live ``close-by" in the attractor sense, but have smaller energy.
\end{itemize}
In Fig.~\ref{fig:energyoscillons} we compare the energy of the single-frequency ansatz with the energy obtained from the quasi-breather formalism, which contains contributions from higher harmonics. We see that the difference is small.

\begin{figure}[h]
    \centering
    \includegraphics[width=\textwidth]{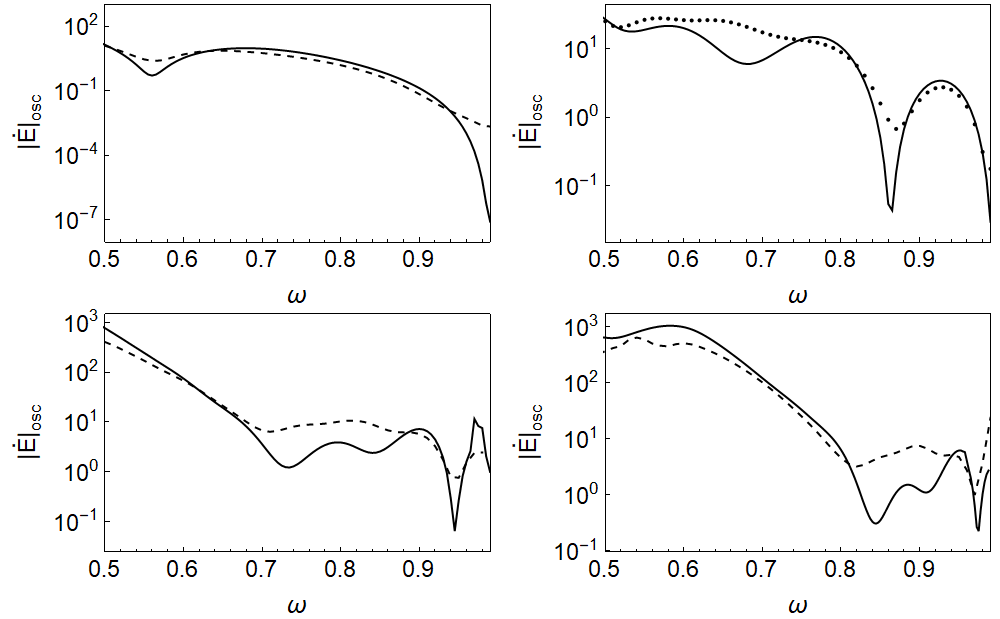}
    \caption{Radiated power obtained analytically using the quasi-breather formalism (solid) and numerically using the prescription described in the main text (dashed) for the oscillon configurations of the sine-Gordon model, for different number of nodes in the spatial profile: zero node (top left), one node (top right), two node (bottom left) and three node (bottom right) solutions. The two methods provide independent estimates at fixed $\omega$. Although they agree qualitatively well, there are quantitative differences. The actual time evolution of an oscillon follows the overall estimate quite closely as it evolves adiabatically, increasing $\omega$.}
    \label{fig:radiationoscillons}
\end{figure}

In our numerical experiment we observe that an initial condition of the form of the single-frequency ansatz \eqref{eq:singlew} generically latches onto an oscillon configuration with corresponding fundamental frequency set by Eq.~\eqref{eq:profile}, but with non-negligible contributions from higher harmonics. Notice the nomenclature here; an oscillon configuration is not equivalent to our definition of a state: either the configuration decays quickly or it changes its features too much throughout its lifetime. These configurations do, however, define a continuously connected space of oscillon solutions, whose properties we can measure (soon after initializing)\footnote{In practice we take measurements by initializing with the single-frequency ansatz, waiting a few periods for the system to relax and finally measuring observables over one or two oscillation periods.}. These solutions have the full harmonic content sourced by the single-frequency ansatz and can thus be compared to the analytically constructed quasi-breathers of Sec.~\ref{sec:model}. Of particular interest is the energy that the oscillon solutions radiate, which we can obtain through the expression
\beq
|\dot{E}_{osc}| = 4 \pi R^2 \langle T_{0r}\rangle|_{r = R} = 4 \pi R^2 \left \langle \dot{\phi} \partial_r \phi \right \rangle|_{r = R}\, ,
\eeq
where brackets indicate time averages and the measurement should be taken at a radius $R$ that is away from the oscillon bulk. We obtain numerical data by initializing with the single-frequency ansatz at different values of $\omega$, letting the field relax to an oscillon solution for a fixed amount of time, and then measuring $T_{0r}$ away from the oscillon bulk for three oscillation periods. $|\dot{E}_{osc}|$ for each $\omega$ is obtained by averaging the measured values of $T_{0r}$. This procedure defines another way to estimate the oscillon properties at fixed $\omega$, independent from the quasi-breather formalism and rooted entirely in numerics. In Fig.~\ref{fig:radiationoscillons} we compare the energy radiated by the quasi-breathers constructed in Section~\ref{sec:model} and the oscillon solutions in our experiment.  We see that the oscillon solutions agree qualitatively well with the quasi-breather formalism, although the latter often underpredicts the amount of energy that the oscillon solution radiates. This is on one hand due to the fact that the we are forced to include a finite number of harmonics in the quasi-breather calculation (a limitation that the numerical measurement doesn't have), and on the other hand because of the uncertainty in our prescription for numerically measuring the properties of the oscillon solutions at fixed $\omega$.

As seen before in Section~\ref{sec:model}, the radiated energy is highly suppressed for specific values of the frequency $\omega$.  Assuming that the configurations are  able to adiabatically follow the ``instantaneous" space of oscillon solutions, we can predict  where proper oscillon states should exist: these are those configurations in the space of solutions that are able to retain their global properties over a substantial amount of time. 
For example, there are clear plateaus when investigating the evolution of the system through $\omega(t)$ using Eq.~\eqref{eq:omegavst} (see Fig.~\ref{fig:omegavstime}), during which the configuration is able to retain its global properties. While the oscillon configuration lives on the plateau it has a well defined mass and binding energy. Note that these plateaus do not necessarily exist during the evolution of the system, as this reasoning is based on the assumption that the oscillon configurations are adiabatically connected. However, it gives a hint as to where we might find oscillon states: oscillon configurations that retain their global properties over a long period of time. In Section~\ref{sec:definition} we make these notions more concrete by providing a definition of an oscillon state.

\subsubsection{Definition of Oscillon States and their Lifetime}
\label{sec:definition}
We define an oscillon state as an oscillon configuration that during its lifetime satisfies the following criteria:
\begin{enumerate}
    \item The fundamental frequency (or binding energy per particle) of the solution doesn't change by more than $10\%$.
    \item The state has a well defined mass, meaning that it doesn't vary by more than $20\%$.
    \item If the spatial profile contains a node, it must retain it.
    \item The overall lifetime has to be at least $50\%$ of the  ground state (lowest energy oscillon state).
\end{enumerate}
With these definitions we  look for oscillon states ``experimentally'', meaning through a series of simulations. If the dynamical evolution of the system were fully adiabatic we could find all the states immediately by numerically solving Eq.~\eqref{eq:omegavst}, essentially combining the data in Figs.~\ref{fig:energyoscillons} and \ref{fig:radiationoscillons}. This process gives clear predictions as to where the oscillon should be able to obey our definition of a state. 
While our simulations confirm that this happens for the nodeless oscillon configurations, it does not occur for oscillons that contain a node in their spatial profile. These solutions eventually shed their node and leave the adiabatic curve. We thus consider the adiabatic predictions as the most probable locations in parameter space where well-defined oscillon states exist and look for them through a series of simulations, initialized with different profiles derived from Eq.~\eqref{eq:profile}.

\subsection{Results}
Here we present the main results of this paper. Using the definition of a state given in Section~\ref{sec:definition} we are able to extract a discrete spectrum of oscillon states with different masses. By increasing the number of nodes in the spatial profile a seemingly boundless spectrum of states emerges. We explicitly checked that states can be found with up to 3 spatial nodes, and have no a priori reason to think that this process should stop (although stability issues might require the initial conditions to be highly fine-tuned).

\begin{figure}[h]
    \centering
    \includegraphics[width=\textwidth]{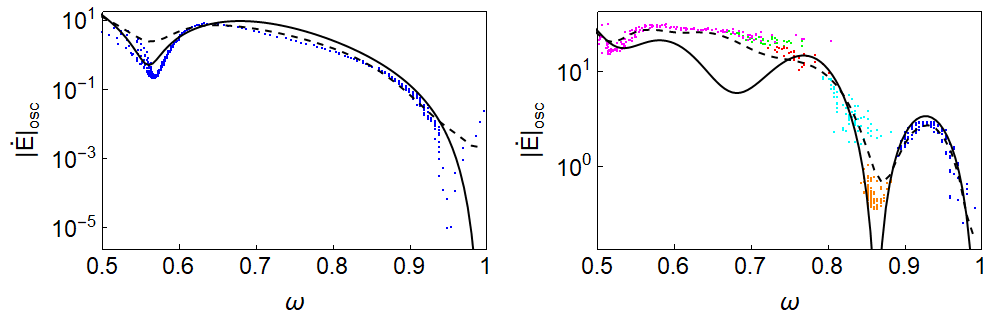}
    \caption{The radiation emitted by the oscillon versus its frequency $\omega$ extracted from the full numerical evolutions starting with different initial conditions, represented by different colors. The black curves are the analytic (solid) and numerical (dahsed) estimates shown in Fig.~\ref{fig:radiationoscillons}. \textit{ Left:} The nodeless case. A single initialization traces out the whole range of $\omega$. \textit{Right:} The case with one node. In this case, the node is lost past some $\omega$ and multiple initial conditions (different colours) are needed to cover the entire range.
    }
    \label{fig:adiabaticcolor}
\end{figure}

\begin{figure}
    \centering
    \includegraphics[width = \textwidth]{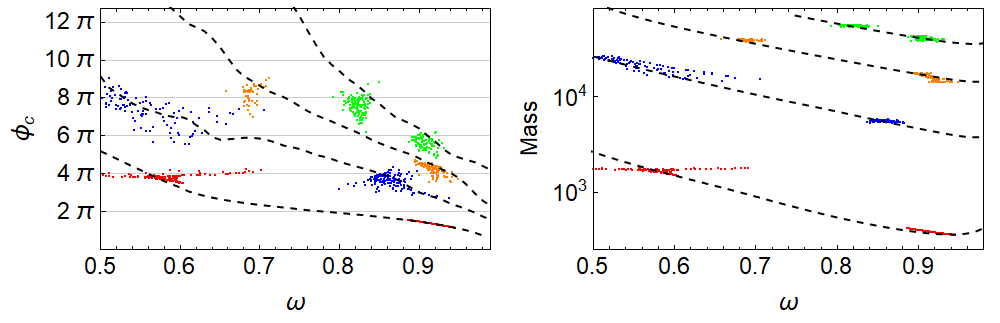}
    \caption{The clustering of oscillon-like configurations in parameter space. On the left panel we plot the maximal central field amplitude (measured over one period) while on the right panel we show the mass of the states, defined as the energy that remains localized within a radius that initially contains $90\%$ of the total energy. It is clear that the field clusters around points which we define as states in this paper. Different clusterings correspond to different initializations while clusterings with the same color correspond to states with the same number of nodes: 0 (red), 1 (blue), 2 (orange) and 3 (green). The black dashed line in the correspond to the predictions of the oscillon configurations obtained from numerics at fixed $\omega$.}
    \label{fig:states}
\end{figure}

As mentioned previously, there is a clear difference in the evolution of the solutions with- and without nodes. Starting from the single-frequency ansatz, the nodeless solution is able to adiabatically probe all the oscillon configurations\footnote{This is strictly speaking only true when starting from the single-frequency ansatz. More general initial conditions can excite instabilities and leave the adiabatic curve.}, up until its "energetic death"\footnote{This term was first coined in \cite{Cyncynates:2021rtf} to indicate the moment where the oscillon has to decay as adiabatic evolution requires an increase in energy of the solution, which is not available due to energy conservation.}.  Solutions with nodes typically excite instabilities that drive them away from this type of adiabatic evolution. We will comment on the nature of these instabilities later. The distinction in behavior is made clear in Fig.~\ref{fig:adiabaticcolor} where we plot (colored scatter) the radiation that the oscillon emits during its entire lifetime, until either no localized energy remains or the field loses its node. Different colors represent different simulations initialized with the single-frequency ansatz. It is clear that the nodeless oscillon is able to evolve adiabatically from $\omega \sim 0.5$ until $\omega \sim 1$ where it decays. An oscillon with  nodes (in Fig.~\ref{fig:adiabaticcolor} we only show the case with one node, but similar conclusions hold for solutions with more nodes) generally loses its node(s) before it can transition to $\omega \sim 1$. We need more than one initialization to map the entire space of oscillon solutions. Since there is no way of knowing in what region of parameter space oscillons with nodes evolve adiabatically, we are forced to look for states through a process of trial and error. However, the information gathered so far allows us to do so in a systematic way:
\begin{enumerate}
    \item Using the predictions for the adiabatic evolution of the system we identify potential oscillon states.
    \item Once identified, we check through numerical simulation whether the oscillon is able to survive long enough to qualify as a state (starting from the single-frequency ansatz).
    \item We explicitly check if the evolution of the oscillon follows our adiabatic prediction.
\end{enumerate}
If the oscillon survives these three steps of scrutiny we identify it as a proper state and take measurements of its properties by averaging over its lifetime. The process is less complicated for the nodeless oscillons as they evolve adiabatically. This predicts the existence of two states which were previously found in Ref.~\cite{Hormuzdiar:1999uz}.
In Fig.~\ref{fig:states} it becomes evident why we classify certain oscillons as states. The different clusterings in parameter space indicate that there are certain preferred oscillon configurations which we identify as states. Each clustering corresponds to a different initialization of the field with the single-frequency ansatz. We gave clusterings belonging to profiles with the same number of nodes the same color. The datapoints in Table~\ref{tab1} are computed by averaging over the different clusters. Note that this is true in particular for the fundamental frequency $\omega$, which is computed by measuring the zero-crossings of the field at the origin. Since $\omega$ evolves somewhat during the evolution of the state, the values given in Table~\ref{tab1} do not necessarily correspond to those of the single-frequency initialization. In fact, to observe the state it is better to initialize with an $\omega$ that is somewhat below the value given in Table~\ref{tab1}. Finally, masses are defined as the energy that remains localized within a radius that initially contains $90\%$ of the total energy. We will now discuss some characteristics of the different states in more detail.

\subsubsection{Fundamental Oscillon States}

There exist two nodeless oscillon states in the sine-Gordon model in three spatial dimensions. We explicitly confirmed their existence and characteristics. Their space- and time- profiles are shown in Fig.~\ref{fig:noNodes}. The heavier state has a fundamental frequency of $\omega \sim 0.58$ and has about four times the mass of the lighter state with $\omega \sim 0.92$. The lighter state turns out to be the lightest state we find in our oscillon spectroscopy of the sine-Gordon model (also including  oscillons with nodes). We therefore refer to it as the ground state. Interestingly, the lifetime of the lower frequency (heavier) state is actually larger than that of the ground state.

What seems typical for the nodeless oscillons is that once the field settles into an oscillon configuration, it can evolve adiabatically very efficiently, increasing its frequency $\omega$. In this way, the scalar field can transition through the two nodeless states if initialized with the right initial conditions, effectively having a coherent lifetime that is the sum of that of the two individual states. In particular, this is what happens when initializing with the single-frequency ansatz (see Fig.~\ref{fig:adiabaticcolor}) near $\omega \sim 0.58$, but does not necessarily happen when starting with more generic initial conditions. We will comment on these types of transitions in Section~\ref{sec:transitions}. Eventually, when the oscillon reaches a frequency of $\omega \eqsim 0.93$, it decays. At this point the oscillon needs to increase its energy in order to continue to increase $\omega$. As this is not possible, the state quickly decays. This type of decay was coined ``energetic death'' in \cite{Cyncynates:2021rtf} and seems to be generic for oscillons in potentials that can be written as a power series. As mentioned previously, the decay of the states with nodes is of a different nature.

The two nodeless states act as very strong attractors of the sine-Gordon model. We observed that, starting from a generic Gaussian field profile, the field quickly settles into one of the nodeless states (which of the two is somewhat dependent on initial conditions). In this case the adiabatic transition through the two states is somewhat less efficient, since the oscillon configurations are perturbed to a larger degree by the initial conditions. In other words, adiabatic evolution can stop even for nodeless states before ``energetic death''.

\subsubsection{Excited Oscillon States}
\begin{figure}[h]
    \centering
    \includegraphics[width = \textwidth]{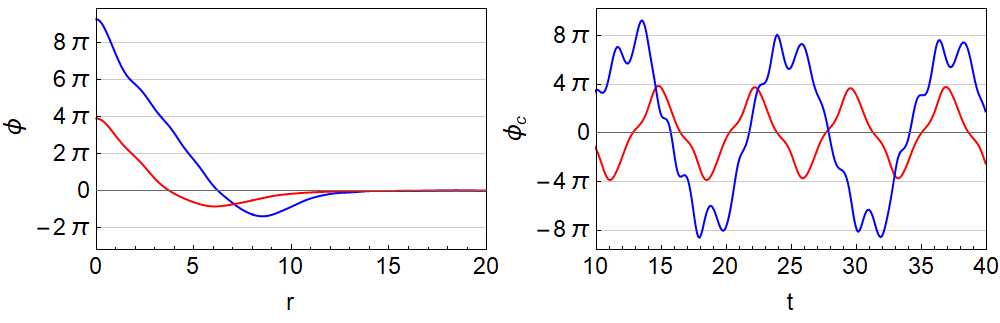}
    \caption{The spatial and temporal profiles for the oscillon states found with a single node in their spatial profiles. In the spatial profile (left) we show the field profiles of the oscillon at maximum amplitude, while in the temporal profile (right) we show the amplitude of the field at the origin as the state evolves.}
    \label{fig:n1states}
\end{figure}

\begin{figure}[h]
    \centering
    \includegraphics[width = \textwidth]{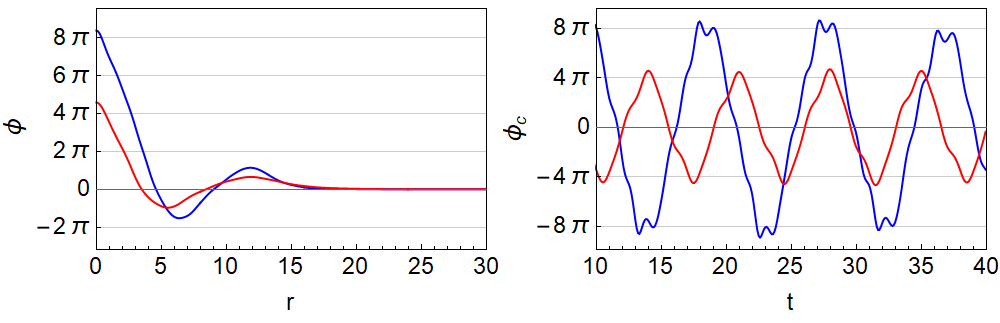}
    \caption{The spatial and temporal profiles for the oscillon states found with two nodes in their spatial profiles.}
    \label{fig:n2states}
\end{figure}

\begin{figure}[h]
    \centering
    \includegraphics[width = \textwidth]{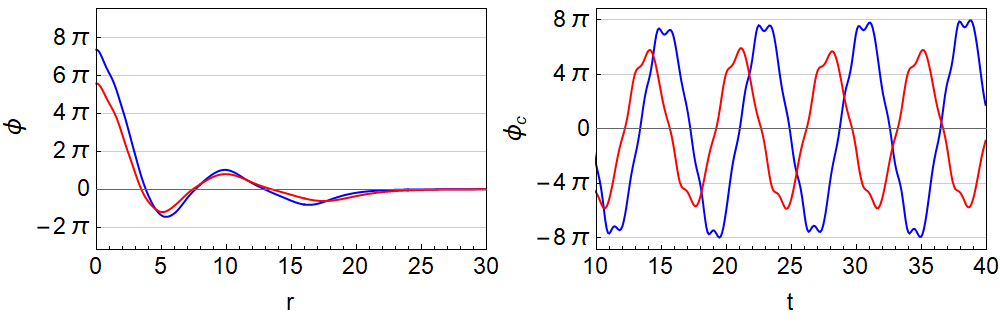}
    \caption{The spatial and temporal profiles for the oscillon states found with three nodes in their spatial profiles.}
    \label{fig:n3states}
\end{figure}

Here we present the oscillon states we found for oscillons that contain nodes in their spatial profile. Interestingly, we are able to identify two states for each number of nodes, replicating ``accidentally'' the pattern of the nodeless states. In Figs.~\ref{fig:n1states},~\ref{fig:n2states} and~\ref{fig:n3states} we present their space- and time- profiles. Interestingly, as the field oscillates around $0$ the nodes stay approximately in the same location. Furthermore, even though we start from a single-frequency ansatz, the states get large contributions from higher harmonics. In this sense these states are also attractors of the model, although we didn't observe them starting from a Gaussian profile, and instead needed to initialize with profiles that contained the correct number of nodes. The less attractive nature of these oscillons ultimately ties in with the fact that states with nodes contain instabilities that force them to decay before energetic death.

Initializing with the single-frequency ansatz we see that the oscillon temporarily is able to evolve following the prediction of Eq.~\eqref{eq:omegavst}. Naively, we would expect the oscillon to then transition through the two states and eventually decay due to energetic death, similar to the nodeless case. However, this type of transition was never observed, and a different type of instability, which forces the oscillon to lose its node(s) takes over. We suspect the origin to be the following: as the scalar waves from the bulk of the oscillon move through the nodes towards spatial infinity, the location of the node has to oscillate. This eventually forces the oscillon to shed its node as it strays too far from adiabatic evolution. We observe that the shedding of the node(s) is a violent event in which a lot of scalar radiation is emitted. Depending on initial conditions, the field sometimes decays into one of the nodeless states.

One might expect that, starting with initial conditions that are ``closer'' to the true oscillon solution (so a solution that contains content from higher harmonics instead of the single-frequency ansatz), one could have a delayed decay of the states as the initial perturbation of the system is smaller. To test this hypothesis we constructed oscillon configurations containing contributions from higher harmonics using the quasi-breather formalism, and used this as initial conditions instead of the single-frequency ansatz. Doing this, we observed slightly altered lifetimes, which we have taken into account by adding uncertainties to our inferred lifetimes.

Table~\ref{tab1} contains the most important characteristics of the long-lived oscillon states with up to three nodes. We see that, while the energy is monotonically increasing, as we increase the number of nodes, the same is not true for the lifetime. The lifetime is of the same order for most long-lived states that we found, ranging from $300 m^{-1}$ to $900 m^{-1}$. We see that the two-node oscillons are less long-lived than both single-node and three-node ones, indicating that it is at least plausible that long-lived states with more nodes can be found. On the other hand, the rest mass increases between the most stable $0$-node and the $3$-node oscillons, ranging from $M=3.8\times 10^{2}$ to $M=5.6\times 10^4$ in units of $f^2/m$. 

{The product $M\, \tau$ is also quite intriguing. It is natural to identify $1/\tau$ as the decay width and view these states as resonances, as usual in particle physics. The outcome is that they correspond to very narrow resonances. Then $M\,\tau$ measures the narrowness of the state (the larger the narrower). It is quite large, and similar for all long-lived oscillon states, 
$M \tau \simeq  \big( {\cal O}(10^6) - {\cal O}(10^7) \big) (f/m)^2$.
It even increases mildly with the number of nodes. (As we discuss in Sec.~\ref{sec:transitions}, away from the spherical ansatz, the lifetime is reduced by about 1 order of magnitude for oscillons with nodes).}
Finally, it is worth noting that the maximum central amplitude of the field profile seems to usually be equal to a multiple of $2 \pi$ which correspond to the minima of the sine-Gordon potential.

\section{Transitions}
\label{sec:transitions}
Transitions between oscillon states can be classified in two ways. First, a state can transition when it excites instabilities, perturbing the oscillon solution and forcing it to decay to one or more decay products. Since these instabilities can be aspherical, the problem needs to be studied in three dimensions. The second type of transition has already been alluded to in the text: an adiabatic transition between oscillon states. We reemphasize this point before moving on to the three dimensional question.

\subsubsection{Adiabatic Transitions}
\begin{figure}[h]
    \centering
    \includegraphics{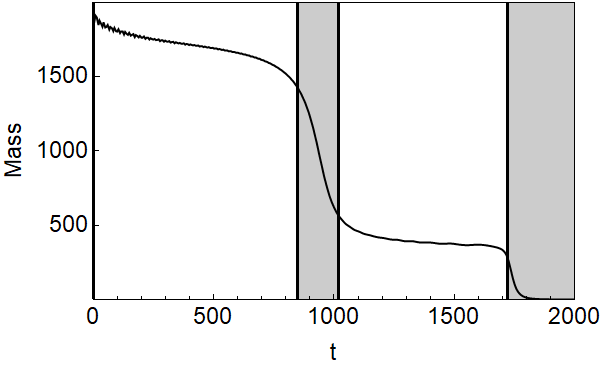}
    \caption{The adiabatic transition between the two nodeless  oscillon states of the sine-Gordon model. Starting from the single-frequency ansatz, the field first settles into the low $\omega$ state (right white region) before transitioning (middle gray region) to the groundstate. Eventually, the groundstate decays quickly due to energetic death.}
    \label{fig:adiabatictransition}
\end{figure}

As noted before, the nodeless oscillon solutions typically evolve adiabatically until they decay due to energetic death. The nodeless oscillon state with $\omega \sim 0.58$ is therefore able to transition to the groundstate at $\omega \sim 0.92$ as energy radiates away from the oscillon bulk. The transition can be seen clearly in Fig.~\ref{fig:adiabatictransition}, where we plot the mass of the oscillon solution over time (defined as the energy within the radius that initially contains $90\%$ of the total energy in the field) for a field configuration initialized with the single-frequency ansatz at $\omega = 0.55$. The field transitions through the two states where its mass is approximately constant (white regions), separated by a fast adiabatic transition (middle gray shaded region). Finally, it decays through energetic death (right gray shaded region).
A transition as in Fig.~\ref{fig:adiabatictransition} always happens when initializing with the single-frequency ansatz, but for completeness we report that, starting from more generic initial conditions (as e.g. a Gaussian), we observed a slightly different evolution with an earlier onset of decay. The transition through the second state was less complete in these cases due to the larger initial perturbation to the true oscillon solution.

\subsubsection{Transitions in 3D}
\begin{figure}[h!]
    \centering
    \includegraphics[width=.9\textwidth]{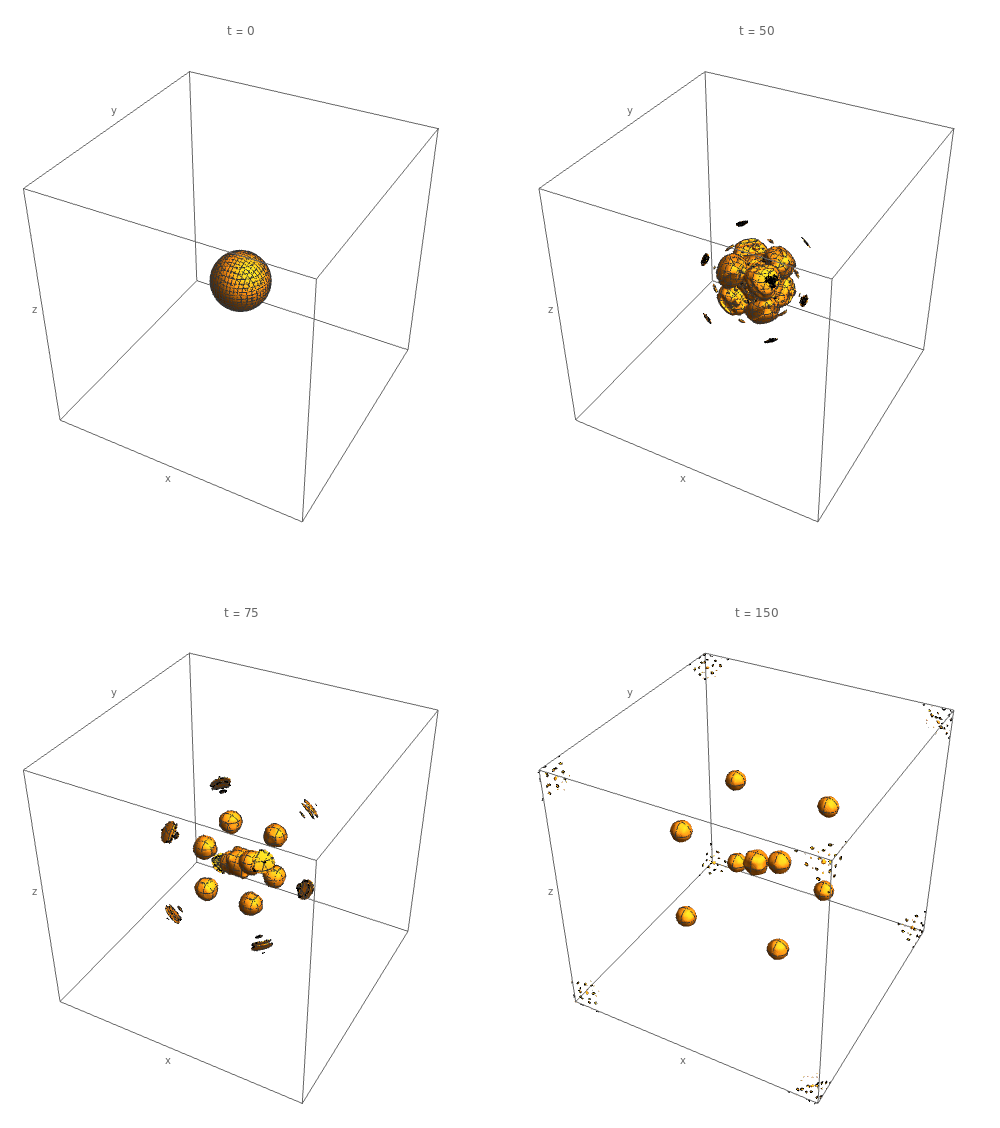}
    \caption{Snapshots of the lattice simulations starting with the oscillon state with one spatial node and frequency  $\omega \sim 0.56$. Due to the fact that numerical perturbations get amplified, the state breaks up into $9$ nodeless oscillons (one with $\omega \sim 0.58$ at the center and eight groundstates with $\omega \sim 0.92$, all oscillating out-of-phase with the central oscillon, moving outwards along the diagonals of the simulation box). The timescale of decay is about one order of magnitude shorter than those predicted by the spherical simulations. Contours are drawn around volumes that have energy density $20$ times the average density in the box ($\rho = 20 \bar{\rho}$).}
 \label{fig:n1lowlattice}
\end{figure}

\begin{figure}[h!]
    \centering    \includegraphics[width=.9\textwidth]{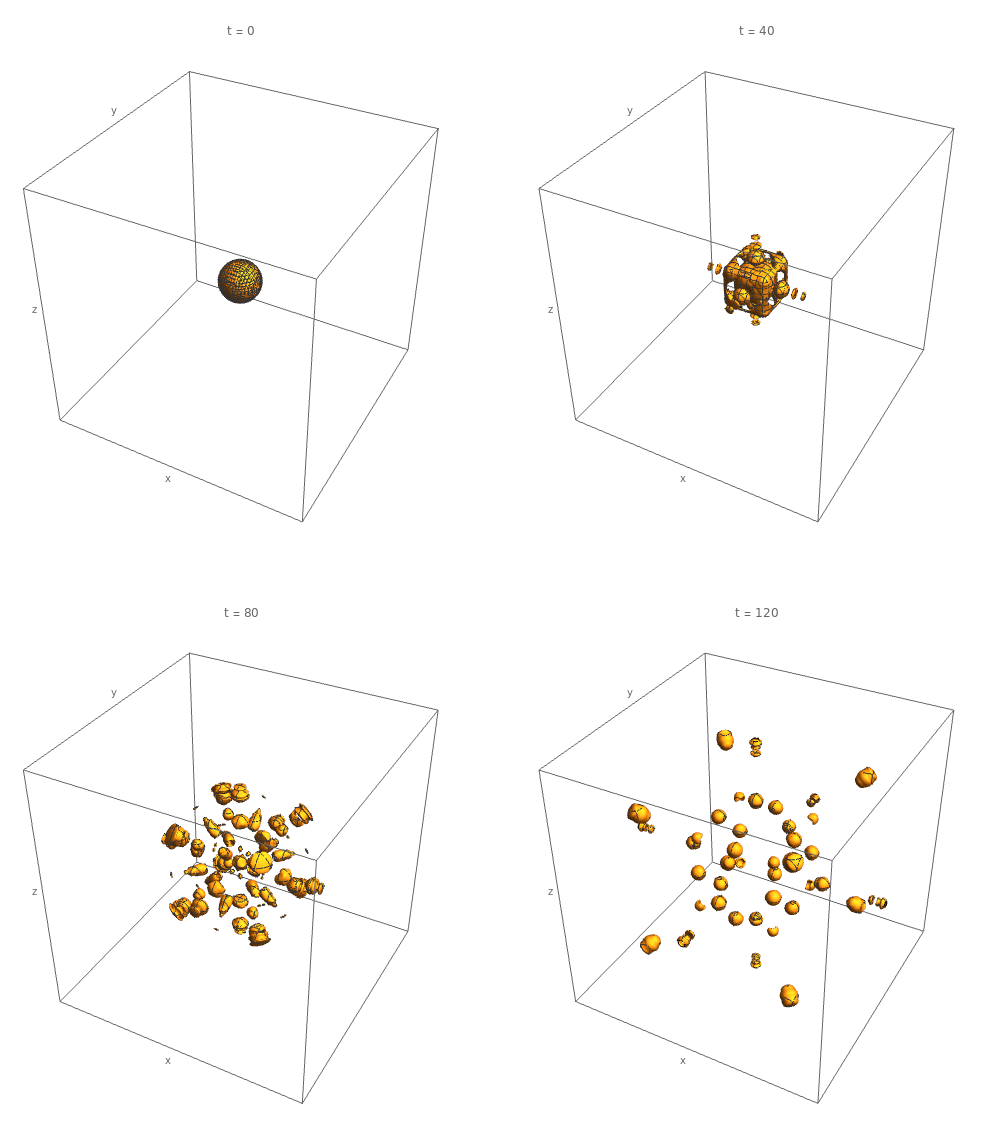}
    \caption{Snapshots of the lattice simulations starting with the oscillon state with three spatial nodes and frequency  $\omega \sim 0.84$. Similar to Fig.~\ref{fig:n1lowlattice}, we see a breakup into nodeless oscillons, but due to the large amount of energy in the initial conditions we end up with many more final oscillons: a spectroscopic signature of   oscillon states with several nodes. Contours are drawn around volumes with $\rho = 50 \bar{\rho}$.}
    \label{fig:n3lowlattice}
\end{figure}

In the previous sections we computed the lifetime of excited sine-Gordon oscillons by restricting the evolution to spherically symmetry. However, localized configurations in three spatial dimensions can decay aspherically, in some cases leading to the formation of groups of localized objects, see e.g. Ref.~\cite{Kinach:2022jdx}. Even starting from spherical initial conditions it's still a possibility that quantum mechancical perturbations get parametrically amplified due to coupling with the spherical background. There is no good reason to expect that this amplification is necessarily smaller for aspherical modes than for spherical ones, and the lifetimes given in Table~\ref{tab1} could be altered significantly in 3+1 dimensions. To investigate this we performed three-dimensional lattice simulations of the states we found in spherical symmetry. We again took as initial conditions the spherical single-frequency ansatz of the various states found in spherical symmetry, which were then perturbed with small fluctuations. For the nodeless states, we see no significant difference in lifetime. However, it seems that oscillon states with nodes are susceptible to aspherical transitions to one or more of the nodeless states. These transitions persist, even when no fluctuations are added to the initial conditions. 
This indicates that numerical perturbations alone can be amplified and trigger the transitions. In Figs.~\ref{fig:n1lowlattice} and ~\ref{fig:n3lowlattice} we show some snapshots of these simulations. One can note that the perturbation is numerical as the breakup follows the symmetry of the underlying cubic lattice. Beyond the limitations of inherent numerical noise, there is reason to believe that the decay itself is a physical effect. 

{We can obtain some heuristic understanding of the fragmentation process shown in Figs.~\ref{fig:n1lowlattice} and \ref{fig:n3lowlattice} as follows.
Considering  the limit where the amplitude at the center is large $\Phi_0\gg1$, 
the equation of motion of a small perturbation living atop of the oscillon state in Fourier space  is
\beq\label{stab}
\Ddot{\delta \phi_k} + \left(k^2 + \cos(\phi_{osc}(t,r))\right) \delta \phi_k = 0
\eeq
with $\phi_{osc}(t,r)$ being the background oscillon. 
For $\Phi_0\gg1$, the time dependent effective mass has a large harmonic composition. At the center we can approximate $\cos(\phi_{osc}(t,0)) \simeq \cos(\Phi_0 \cos(\omega t)) $, which contains frequencies up to $\sim\Phi_0 \omega$. It seems feasible that high $k$ modes can be resonantly excited. 

Let us assume that in the large $\Phi_0$ limit the oscillon width $R_{osc}$ is fixed, as Figs.~\ref{fig:n1states}, \ref{fig:n2states}, \ref{fig:n3states} suggest. Then, a separation of scales appears $\Phi_0 \omega \gg 1/R_{osc}$, and one can to see whether modes with $k\gg 1/R_{osc}$ are resonant by switching to the homogeneous problem
\beq
\Ddot{\delta \phi_k} + \left(k^2 + \cos(\Phi_0 \cos(\omega t))\right) \delta \phi_k = 0
\label{eq:unstablefloquet}
\eeq
of which the instability bands can be found using Floquet theory. Indeed, Eq.~\eqref{eq:unstablefloquet} leads to unstable bands at large values of $k$, up to a maximal value that scales approximately as $k_{max} \propto \omega \Phi_0$. 

The conclusion of this is that we can expect instabilities for modes in the range $R_{osc}^{-1} \lesssim k \lesssim \Phi_0 m$. On the other hand, the disctrete lattice inevitably introduces ``noise'' with the lattice symmetry that  sources these unstable modes with same symmetry.  

This gives a qualitative picture of the observed fragmentation effect. States with the amplitude $\Phi_0 \sim 8\pi$ can be expected to have a number of unstable modes. 
Indeed, the states shown in Fig.~\ref{fig:n1lowlattice} and Fig.~\ref{fig:n3lowlattice} both have an initial amplitude $\sim 8\pi$ and develop several lobes which then evolve into oscillons (accordingly to energy conservation of course).
For states with lower amplitude the argument loses validity. Indeed, we find that of the two states with $\Phi_0\sim4\pi$, (the heavy $n=0$ state and the light $n=1$)  only the $n=1$ state (the heavier of the two) fragments. 

}

Following this argument, and using the hints obtained from the lattice simulations we performed, it seems that the true lifetimes of many of the states found in previous sections are in reality significantly shorter, {by about 1 order of magnitude.} These considerations also highlight the incredible properties of the nodeless state living around $\omega \sim 0.58$. It is the longest lived state that we found, reaches a field amplitude comparable to some of the states with nodes, is highly relativistic as a bound state with $\omega \approx 0.58m$, and finally, is stable against aspherical decay even though it has about four times the mass of the groundstate.

Our lattice simulations reveal new interesting features that emerge in three dimensions. Namely, the different states should decay through a rich spectrum of transitions to the more stable states. This is a different type of ``spectroscopic'' feature which can be used to characterize the states. This is a topic we plan to address in future work. Furthermore, although we observe aspherical decay of the oscillon states  with nodes in the sine-Gordon equation, following our reasoning in the previous paragraph there is no reason to think this is bound to happen in other models. There should in principle exist systems where  states with nodes are just as stable as nodeless states; a question we will tackle in the future.

\section{Conclusions}
\label{sec:conclusions}

We found radically new spherically symmetric oscillon solutions in the three dimensional sine-Gordon equation. Their characteristic is the existence of nodes in their spatial profile, which they keep throughout their lifetime. They 
have significantly larger energy than their fundamental (nodeless) counterparts, in particular solutions with just one node exhibit one order of magnitude larger energy.
We provided a semi-analytic construction by using the quasi-breather formalism, taking into account the non-perturbative presence of higher harmonics. This construction leads to a two-parameter family of solutions, discrete in the number of nodes and continuous in the frequency. However, the radiating tails of the quasi-breathers lead to the existence of long-lived oscillons, only around certain frequencies, with lifetimes reaching ${\cal O}(10^{3})\, m^{-1}$.

Despite the vast difference in the mass (rest energy), the lifetime of excited oscillons is similar to their single field counterparts, when computed under spherical symmetry. It is interesting that the lifetime is not monotonic; oscillons with three nodes exhibit longer lifetimes than their two-node or even fundamental counterparts. 
By considering oscillons as resonances in the particle spectrum of the theory, we can define their  narrowness of as the product of the mass and lifetime. Surprisingly, this grows mildly with the number of nodes, being ${\cal O}(1)$ for nodeless oscillons and ${\cal O}(10)$ for oscillons with three nodes.

When perturbed outside the spherical ansatz, using a full three-dimensional simulation, the excited oscillons exhibit a significantly smaller lifetime of ${\cal O}(30)\, m^{-1}$. 
Interestingly, the decay does not lead to an incoherent bath of radiation, but instead leads to the formation of a number of fundamental oscillons. The possible existence of a ``selection rules'' for the decay of excited multi-node oscillons is beyond the scope of the present work, but does present an intriguing challenge to which we will return in the future. 

{Our results add more remarkable properties to the heavy oscillon without nodes of the theory. Its low frequency indicates that it can be regarded as a relativistic bound state. Moreover, in our 3D lattice simulations we find it to be stable against fragmentation or decay anisotropically. This happens despite having about 4 times the mass of the lowest oscillon, and despite the field excursion being quite large, $\pm 4\pi f$, at the center. }

Finally, we must note that the sine-Gordon model is certainly special. 
The potential has maxima, and with sizeable $V''_{\rm max}= - V''_{\rm min}$. 
It is natural to ask how many features (if any) of the  current analysis carry on to more general potentials. 
We will explore the change in the oscillon spectrum from small or large deviations from the sine-Gordon model in a future publication.
\\\\

\section*{Acknowledgements}
	{We thank  Mark Hertzberg, Alex Pomarol  and Sergey Sibiryakov for useful discussions.} The research leading to these results has received funding from the Spanish Ministry of Science and Innovation (PID2020-115845GB-I00/AEI/10.13039/501100011033). IFAE is partially funded by the CERCA program of the Generalitat de Catalunya. EIS  acknowledges support of a fellowship from ``la Caixa'' Foundation (ID 100010434) and from the European Union’s Horizon 2020 research and innovation programme under the Marie Skłodowska-Curie grant agreement No 847648. The fellowship code is LCF/BQ/PI20/11760021. The research leading to these results has
    received funding from the ESF under the program
    Ayudas predoctorales of the Ministerio de Ciencia e
    Innovación PRE2020-094420.
	
\appendix

\section{Multi-frequency oscillon solution}
\label{app:multifrequency}

While the single frequency Ansatz is sufficient to describe the low-energy oscillon state(s), we need to go beyond it in order to better capture the structure of excited  oscillons; meaning oscillons with nodes that have higher energy. We follow the method put forth in Ref.~\cite{Cyncynates:2021rtf}, which is tailored to the study of potentials that can be written in the form $V(\phi) \sim \sum_n {\cal V}_n \left [1-\cos(n\phi)\right ]$ (note that we have dropped the axion decay constant $f$ for simplicity).  The sine-Gordon model represents the simplest case of this family of potentials.
We do not attempt to repeat the method of Ref.~\cite{Cyncynates:2021rtf} in its full generality, but rather to present the basic steps, in order to make the current work self contained.  

The first step in constructing a viable oscillon solution is extending the single frequency Ansatz to contain multiple frequencies. Due to the symmetry of the potential, the oscillon only contains odd multiples of the fundamental frequency (odd harmonics).
\beq
\phi_{sin}(r,t) = \sum_n \Phi_n (r,\omega) \sin (n \omega t) \, .
\label{eq:sinseries}
\eeq

However, we know that the oscillon is not a completely stable configuration and as such it contains small radiating tails, which explain both its longevity and its slow evolution and ultimate decay. In order to capture the  radiating tails, we need to add a cosine series to Eq.~\eqref{eq:sinseries}, in particular
\beq
\phi_{cos}(r,t) = \sum_{n\omega>m} { c}_n (r,\omega) \cos (n \omega t) \, ,
\label{eq:sinseries}
\eeq
where the cosine series only contains radiative modes with $n\omega>m$. Finally, our localized slowly radiating solution can be written as
\beq
\phi_{\rm full} \simeq \phi_{sin}(r,t)+\phi_{cos}(r,t) \, .
\label{eq:phiosc}
\eeq
We see that (e.g. Fig.~\ref{fig:node0analytics}), even for radiative modes with $n\omega>m$, the behavior close to the core of the oscillon and in the radiative tail can be very different.
We insert the expansion of Eq.~\eqref{eq:phiosc} into the equation of motion \eqref{eq:SGeom} 
and use the Jacobi-Anger expansion 
\beq
e^{i \alpha \, \sin b}  = \sum_{k=-\infty}^\infty J_k(\alpha) e^{ikb}\,
\label{eq:jacobianger}
\eeq
where $J_k$ is the Bessel function of the first kind. If we assume that  only the first and third harmonics are important $\phi_{\rm sin}\simeq \Phi_1 \sin(\omega t) + \Phi_3 \sin(3\omega t)$ the equation of motion is written as
\beq
 \Phi_1 \omega^2 \sin(\omega t) + \Phi_3  9 \omega^2 \sin(3\omega t) 
+ {d^2\Phi_1 \over dr^2}
+ {2\over r} {d\Phi_1 \over dr}
+ {d^2\Phi_3 \over dr^2}
+ {2\over r} {d\Phi_3 \over dr}
+ \sin \left [
\Phi_1 \sin(\omega t) + \Phi_3 \cos(3\omega t)
\right ] =0 \, .
\eeq
The last term can be expanded using Eq.~\eqref{eq:jacobianger}
\begin{align}
\begin{split}
\sin(\phi_{sin})  &\simeq 
2 \left [J_0(\Phi_3) J_1(\Phi_1) + J_1(\Phi_3 ) J_2(\Phi_1) + ...\right ] \sin(t)
\\
&+
2 \left [J_0(\Phi_1) J_1(\Phi_3) + (J_0(\Phi_3) - J_2(\Phi_3)) J_3(\Phi_1) + ...
\right ] \sin(3 t)
\, ,
\end{split}
\end{align}
where we kept only a few terms in the Jacobi-Anger expansion for each of the two harmonics. We now see that the equation of motion naturally divides into separate parts, each oscillating with $\sin(t)$ and $\sin(3t)$
\begin{eqnarray}
 {d^2\Phi_1 \over dr^2}
+ {2\over r} {d\Phi_1 \over dr}
+
\Phi_1 \omega^2  -
2 \left [J_0(\Phi_3) J_1(\Phi_1) + J_1(\Phi_3 ) J_2(\Phi_1) + ...\right ] =0
\, ,
\\
{d^2\Phi_3 \over dr^2}
+ {2\over r} {d\Phi_3 \over dr}
+
9 \omega^2\Phi_3   -
2 \left [J_0(\Phi_1) J_1(\Phi_3) + (J_0(\Phi_3) - J_2(\Phi_3)) J_3(\Phi_1) + ...
\right ] =0 \, .
\end{eqnarray}

For practical purposes, we kept a finite amount of Jacobi-Anger terms, making sure that we the truncation is enough to provide the desired accuracy.
It is evident that the above equations do not include the cosine terms and thus cannot provide the proper radiative boundary conditions (outgoing waves) at spatial infinity. We assume that the cosine terms of the expansion in Eq.~\eqref{eq:phiosc} can be treated as a small quantity which does not back-react on the sinusoidal terms. Thus we insert it into the equation of motion and use the Jacobi-Anger expansion, but linearize the resulting potential term, leading to
\beq
{d^2{ c}_3 \over dr^2}
+ {2\over r} {d{ c}_3 \over dr}
+
9\omega^2{ c}_3   -
 \left [
 J_0(\Phi_1)J_0(\Phi_3) - 
  J_1(\Phi_3)J_3(\Phi_1) 
 + ...
\right ] c_3 =0 \, .
\eeq
Since we require a solution with a confined first harmonic we set $\Phi_1(r\to\infty)=0$ and set the other two functions to describe outgoing spherical waves at infinity
$\Phi_3(r)- \sqrt{9\omega^2-1}\, r c_3(r)+ r \Phi'_3(r)=0
$
and $ \sqrt{9\omega^2-1} \, r \Phi_3(r)+ r \Phi'_3(r)+
r c'_3(r)=0
$.
It is evident how this method can be  generalized to include higher harmonics.

\bibliography{SG}

\end{document}